%% file: FisherRaoApproximation-EllipticalFamilyForV4.tex
\documentclass[11pt]{article}

\usepackage{amssymb,amsmath,fullpage,graphicx,url,hyperref}
\usepackage{orcidlink}
\usepackage[boxed]{algorithm2e}

\usepackage{listings}
\input{macrosEl.tex}

\graphicspath{{./}{Fig/.}}

\title{Approximation and bounding techniques\\ for the Fisher-Rao distances between parametric statistical models\footnote{The contents of this paper appeared as a chapter in~\cite{NIELSEN202467}.}}

\author{Frank Nielsen~\orcidlink{0000-0001-5728-0726}\\ \ \\ Sony Computer Science Laboratories Inc\\ Tokyo, Japan}
 
\date{}

\def\Mylabel#1{\label{#1}}

\begin{document}
\maketitle

\begin{center}
Dedicated to the memory of C. R. Rao (1920-2023)
\end{center}

\begin{abstract}
The Fisher-Rao distance between two probability distributions of a finite-dimensional parametric statistical model 
is defined as the Riemannian geodesic distance induced by the Fisher information metric. 
The Fisher-Rao distance is guaranteed by construction to be invariant under diffeomorphisms 
 of both the sample space and the parameter space of the statistical model.
In order to calculate the Fisher-Rao distance in closed-form, one proceeds by (1) eliciting a formula for the Fisher-Rao geodesics,
 and (2) integrating the Fisher-Rao length element along those geodesics. 
These calculations turn out to be difficult tasks for some statistical models 
like the  multivariate elliptical distributions which include the  class of multivariate normal distributions. 
In this work, we consider several numerically robust approximation and bounding techniques for the Fisher-Rao distances:
First, we report  generic upper bounds on the Fisher-Rao distances based on closed-form Fisher-Rao distances of univariate submodels.
Second, we describe several generic approximation schemes depending on whether the Fisher-Rao geodesics or pregeodesics are available in closed-form or not.
In particular, we obtain a generic method to guarantee an arbitrarily small additive error on the approximation provided that Fisher-Rao pregeodesics and tight lower and upper bounds are available. This case applies to the class of multivariate normal distributions.
Third, we consider the case of Fisher metrics being Hessian metrics, and report  generic tight upper bounds on the Fisher-Rao distances as square roots of Jeffreys-Bregman divergences which correspond to the Fisher-Rao energies of dual curves in information geometry. 
Uniparametric and biparametric statistical models always have Fisher Hessian metrics.
Fourth, we consider elliptical distribution families and show how to apply the above approximation and bounding techniques to these models.
We also propose two new distances based either on the Fisher-Rao lengths of  curves serving as proxies of Fisher-Rao geodesics, or based on the Birkhoff/Hilbert projective cone distance.
Finally, we consider an alternative group-theoretic approach for statistical transformation models based on the notion of maximal invariant which yields insights on the structures of the Fisher-Rao distance formula and its degrees of freedom which may be used fruitfully in applications. 
\end{abstract}


\tableofcontents

\section{Introduction}

The notion of dissimilarity~\cite{deza2009encyclopedia} between two probability distributions is essential in statistics~\cite{eguchi2022minimum}, information theory~\cite{arndt2001information}, signal processing~\cite{basseville2013divergence}, and machine learning~\cite{wang2020comprehensive}, among others. 
In general, a dissimilarity $D(P,Q)$ between two elements $P$ and $Q$ is such that $D(P,Q)\geq 0$ with equality holding if and only if $P=Q$.
When the dissimilarity satisfies both the symmetry and the triangle inequality, it is called a metric distance\footnote{Although mathematicians call distances ``metric distances'', the word distance  or metric can be loosely used as a synonym of a dissimilarity measure which may not be a metric distance.}~\cite{gibbs2002choosing}.
When the dissimilarity is smooth, it is called a divergence~\cite{IG-2016} (e.g., the Kullback-Leibler divergence).
Many dissimilarities and  classes of dissimilarities have been studied in the literature, and first principles and properties characterizing those classes have been sought after.
In particular, the class of $f$-divergences~\cite{Csiszar-1967,fdiv-AliSilvey-1966}, class of Bregman divergences~\cite{Bregman-1967}, and the class of integral probability metrics~\cite{muller1997integral} have highlighted key concepts for the notion and properties of statistical dissimilarities.

In statistics, the Mahalanobis distance~\cite{Mahalanobis-1936} (1936) between two normal distributions with same covariance matrix, and the Bhattacharyya distance\footnote{Beware that the Bhattacharyya distance is being loosely called ``distance'' even if it does not satisfy the triangle inequality.}~\cite{bhattacharyya1946measure} (1946) between multinomial/categorical distributions  were first considered for statistical analysis.
In 1945, Rao~\cite{Rao-1945} introduced a generic metric distance for statistical models $\calM=\{P_\theta(x) \st\theta\in\Theta\}$ based on Riemannian geometry which is invariant to diffeomorphisms of both the sample space and the parameter space, and gave some first use cases of this distance for geodesic hypothesis testing and classification of populations.
This Riemannian geodesic distance has been called Rao's distance~\cite{atkinson1981rao,villarroya1993statistical,krzanowski1996rao}, Fisher-Rao distance~\cite{pinele2020fisher}, or Fisher-Rao metric~\cite{liang2019fisher} in the literature.

In general, the Fisher-Rao distance $\rho(\theta_1,\theta_2)$ between two probability distributions $P_{\theta_1}$ 
and $P_{\theta_2}$ of a statistical model $\calM$ is not known in closed-form, even for common models like the
 multivariate normal distributions~\cite{pinele2020fisher,Kobayashi-2023}. 
Thus approximation methods and lower and upper bounds have been studied for widely used statistical models~\cite{EllipticalDistance-CalvoOller-2002,reverter2003computing,pinele2020fisher}.

This paper considers {\em general principles} derived from insights of the underlying Fisher-Rao geometry for approximating and bounding the Fisher-Rao distances.
In particular, we shall consider numerically robust techniques for dealing with the Fisher-Rao distances between multivariate elliptical distributions~\cite{EllipticalDistribution-Kelker-1970}.

We outline the paper with its main contributions as follows:

In \S\ref{sec:FR}, we recall the definition of Riemannian distances (\S\ref{sec:RieDom}) and of Fisher-Rao distances (\S\ref{sec:DefFR}) for  arbitrary (regular) statistical models, and discusses about its computational challenges (\S\ref{sec:comptractability}).
We first explain how the Fisher-Rao distances can be easily calculated  for uniparametric models in~\S\ref{sec:FR1D} 
(Proposition~\ref{prop:fr1d} and Proposition~\ref{prop:FRef1d}) and for  product of stochastically independent models in \S\ref{sec:productmodel} 
(Proposition~\ref{prop:FRindepsto}), and describe a generic technique to get upper bounds on the Fisher-Rao distance in \S\ref{sec:FisherManhattan} (Fisher-Manhattan upper bounds of Proposition~\ref{prop:FisherManhattan}).

Second, we consider generic schemes and algorithms in \S\ref{sec:approxscheme}:
We describe a simple method to approximate the Fisher-Rao lengths of curves in \S\ref{sec:approxFRlength} (Proposition~\ref{prop:approxFRlength}) based on  discretizations and $f$-divergences or length element approximations (Proposition), present an approximation technique relying on a property of metric geodesics in \S\ref{sec:metricspace} 
(Proposition~\ref{prop:frepsgeo}), and approximation algorithms with guaranteed errors when geodesics or pregeodesics are available in closed-form with tight lower and upper bounds
(Algorithm~\ref{algo:multerrorgeo}, Algorithm~\ref{algo:multerrorpregeo}, and Algorithm~\ref{algo:adderror} in \S\ref{sec:adderr}).

In \S\ref{sec:HessianMetric}, we consider Fisher metrics which are  Hessian metrics~\cite{Shima-2007}, and report   upper bounds (Proposition~\ref{prop:ubhessian}) which are square roots of Jeffreys-Bregman divergences on the Fisher-Rao distances and correspond to Fisher-Rao energies of dual curves well-studied in information geometry (Proposition~\ref{prop:JBDRealization}). 
Those  square roots of Jeffreys-Bregman  upper bounds are tight at infinitesimal scales. Examples of Hessian Fisher metrics are Fisher metrics obtained from exponential or mixture families,  2D elliptical distribution families, etc.

In \S\ref{sec:lowerboundiso}, we consider lower bounds which can be obtained by isometric embeddings of Fisher-Rao manifolds into higher-dimensional manifolds: The Fisher-Rao distances distances are preserved whenever the embedded Fisher-Rao submanifolds are totally geodesic. Otherwise, we get lower bounds which are tight at infinitesimal scales.

We then consider Fisher-Rao manifolds of univariate elliptical families in \S\ref{sec:frls} and multivariate elliptical families in \S\ref{sec:elliptical}, and apply our techniques for approximating and bounding the Fisher-Rao distances for these families. 
Elliptical families include the families of generalized Gaussian distributions, $t$-distributions, and Cauchy distributions, among others.
We propose an approximation method based on the Fisher-Rao  lengths of curves obtained by pulling back geodesics of the higher-dimensional cone of symmetric positive-definite matrices, and propose in \S\ref{sec:frhilbertgeo} a novel distance based on Hilbert/Birkhoff cone geometry~\cite{Birkhoff-1957,lemmens2014birkhoff} which is fast to compute since it requires only extreme eigenvalues (Definition~\ref{def:FBdist}) and not the full spectrum.

In \S\ref{sec:maxinvardist}, we discuss another algebraic approach to characterize Fisher-Rao distances of statistical transformation models which are invariant under group actions. We explain how the concept of maximal invariant~\cite{eaton1989group} provides some structural insights on the formula of Fisher-Rao distances, and how this can be used in practice even when the Fisher-Rao distances are not known in closed-form.

Finally, we conclude this work in \S\ref{sec:concl} and discuss some problems for future work.

\section{The Fisher-Rao distance}\Mylabel{sec:FR}

\subsection{Preliminary: Riemannian geometry of domains}\Mylabel{sec:RieDom}

In order to define the Fisher-Rao distance, we first recall the definition of the geodesic distance in Riemannian geometry.
A Riemannian manifold is usually introduced by means of an atlas of local charts equipped with smooth transition maps (see the textbook~\cite{godinho2012introduction} or the concise introduction of~\cite{jost2015mathematical}, \S 5.3.2 p. 142--163).
In this paper, we shall construct Fisher-Rao Riemannian manifolds with atlases consisting of global charts $(\Theta,\theta)$, where $\Theta\subset\bbR^m$ are open domains and $\theta$ global coordinate systems (i.e., Riemannian geometry of domains~\cite{greene2011geometry}).

Consider a symmetric-positive-definite matrix $G(\theta)$ varying smoothly for $\theta\in\Theta$. 
The domain $\Theta$ may be interpreted as a $m$-dimensional manifold $\calM$ equipped with a Riemannian metric bilinear tensor $g$ expressed in the global $\theta$-coordinate system as $G(\theta)=[g_{ij}(\theta)]$. 
Let $\{\partial_1,\ldots,\partial_m\}$ (with $\partial_i=\frac{\partial}{\partial \theta^i}$) form a canonical basis $\calB=\{\partial_1,\ldots,\partial_m\}$ of the vector tangent space $T_p$ at $p\in\calM$. We have $T_p\cong \bbR^m$ and  the $\partial_i's$ can be expressed in the $\theta$-coordinate system by the one hot vector $(\partial_i)^j=\delta_{ij}$ where $\delta_{ij}=1$ iff $i=j$ and $0$ otherwise.  
At a point $p\in\calM$ with coordinates $\theta(p)=(\theta^1(p),\ldots,\theta^m(p))$ (and $\partial_i^j=\theta^j(\partial_i)$), we measure the inner product of  vectors $v_1=\sum_i \partial_i v_1^i$  and $v_2=\sum_j \partial_2 v_2^j$ in the tangent plane $T_p$ as 
$$
g_p(v_1,v_2)=g\left(\sum_i \partial_i v_1^i,\sum_j \partial_2 v_2^j\right)=\sum_{i,j} v_1^i v_2^j\, g(\partial_i\theta,\partial_j \theta)= 
[v_1]_\calB^\top\, \times\, G(\theta)\, \times\, [v_2]_\calB,
$$ 
where $[v_i]_{\calB}$ denote the components of the column vector $v_i$ in basis $\calB$ for $i\in\{1,2\}$, and $g_{ij}=g(\partial_i,\partial_j)$.
The length of a vector $v\in T_p$ is $\|v\|_p=\sqrt{g_p(v,v)}$.

In differential geometry, geodesics $\gamma(t)$ are defined with respect to an affine connection~\cite{godinho2012introduction} $\nabla$ as autoparallel curves\footnote{Parallelism according to an affine connection is defined by a first-order ordinary differential equation. Thus autoparallel curves are curves with tangent vectors parallel yielding a second-order differential equation~\cite{godinho2012introduction}.} satisfying the following second-order ordinary differential equation (ODE):
\begin{equation}\Mylabel{eq:RieOde}    
\forall k\in\{1,\ldots, m\},\quad \ddot\gamma_k+\sum_{i,j} \Gamma_{ij}^k\, \dot\gamma_i\dot\gamma_j=0,
\quad \dot\gamma(t)=\frac{d}{\dt}\gamma(t), \ddot\gamma(t)=\frac{d^2}{\dt^2}\gamma(t)
\end{equation}
where the functions $\Gamma_{ij}^k(\theta)$ are the  $m^3$ Christoffel  symbols of the second kind.\footnote{Christoffel  symbols are not tensors because they do not obey the covariant/contravariant laws of  transformations of tensors~\cite{godinho2012introduction}.
We can transform Christoffel  symbols of the first kind $\Gamma_{ijk}$ into the second kind by $\Gamma_{ij}^k=\sum_m \Gamma_{ijk}g^{mk}$ where $G^{-1}(\theta)=[g^{mk}]$ (inverse Fisher information matrix) and reciprocally transform Christoffel  symbols of the second kind into first kind by  $\Gamma_{ijm}=\sum_k \Gamma_{ij}^k g_{km}$.
}
The ODE of Eq.~\ref{eq:RieOde} can be solved either by fixing initial value conditions (IVC) $\gamma(0)=p_0$ and $\dot\gamma(0)\in T_{p_0}$:
$$
\mathrm{IVC}:\quad \left\{
\begin{array}{l}
\ddot\gamma_k+\sum_{i,j} \Gamma_{ij}^k\, \dot\gamma_i\dot\gamma_j=0\\
\gamma(0), \dot\gamma(0)\in T_{\gamma(0)}
\end{array}\right.
,$$
or by giving the boundary value conditions 
$\gamma(0)$ and $\gamma(1)$:
$$
\mathrm{BVC}:\quad \left\{
\begin{array}{l}
\ddot\gamma_k+\sum_{i,j} \Gamma_{ij}^k\, \dot\gamma_i\dot\gamma_j=0\\
\gamma(0), \gamma(1)
\end{array}\right.
.$$

 Notice that solving the ODEs may be easier  in one case than the other~\cite{MVRao-2005}.
For example, the geodesic ODE was solved with initial value conditions for some particular Riemannian metrics in~\cite{CalvoOller-1991} but not for the boundary value conditions.

In Riemannian geometry, there is a unique torsion-free affine connection induced by $g$ called the Levi-Civita connection $\bar\nabla$ which preserves the metric under  $\bar\nabla$-parallel transport.
The arc length of smooth curve $c$ ($c: [a,b]\rightarrow \calM$, $t\mapsto c(t)$) is defined by integration of the length element 
$\ds(p,\dv)=\|\dv\|_p=\sqrt{g_p(\dv,\dv)}$ along the curve:
$$
\Length(c)=\int_a^b \|\dot c(t)\|_{c(t)} \,\dt,
$$
where $\dot c(t)=\frac{d}{\dt}c(t)$.
The lengths of curves are invariant under monotonic reparameterizations $u=u(t)$, and we may choose the
 canonical arc length parameterization $s=s(t)$ such that $\|\dot c(s)\|_{c(s)}=1$.
In the chart $\theta$, the length elements $\|\dot c(s)\|_{c(s)}$ are expressed as
$$
\|\dot c(s)\|_{c(s)}=\sqrt{\sum_{i,j} g_{ij}(\theta(c(s)))} \, \frac{d}{\ds}(\theta^i\circ c)(s) \, \frac{d}{\ds}(\theta^j\circ c)(s).
$$

It can be shown that the Riemannian geodesics $\gamma_{p_{\theta_0}p_{\theta_1}}(s)$ satisfying Eq.~\ref{eq:RieOde} with boundary conditions
 $p_{\theta_0}=\gamma_{p_{\theta_0}p_{\theta_1}}(0)$ and $p_{\theta_1}=\gamma_{p_{\theta_0}p_{\theta_1}}(1)$ (i.e., $\gamma_{p_{\theta_0}p_{\theta_1}}: [0,1]\rightarrow\calM$) minimize locally the length of curves joining the endpoints $p_{\theta_0}$ and $p_{\theta_1}$.
That is, let $C(p,q)$ denote the set of piecewise smooth curves from $p$ to $q$.
Then the Riemannian distance $\rho(p,q)$ is defined by
\begin{equation}\Mylabel{eq:inflength}
\rho(p,q)=\inf_{c\in C(p,q)} \Length(c).
\end{equation}

\begin{remark}
The geodesic distance can also be calculated from the Riemannian logarithmic map as
$\rho(p,q)=\|\Log_p(q)\|_p$ and the geodesics expressed using the Riemannian exponential map  by $\gamma_{pq}(t)=\Exp_p(t\,\Log(q))$ on Cartan-Hadamard manifolds~\cite{bacak2014convex} (i.e., manifolds with non-positive sectional curvatures).
\end{remark}

The Riemannian distance is a metric distance which satisfies the symmetry and triangle inequality of metric distances.
Furthermore, by the Hopf-Rinow theorem~\cite{godinho2012introduction}, $(\calM,\rho)$ is a complete metric space~\cite{bridson2013metric} and 
there exists a length minimizing geodesic $\gamma_{\theta_1\theta_0}(t)$ connecting any two points $p_{\theta_0}=\gamma_{\theta_0\theta_1}(0)$ and  $p_{\theta_1}=\gamma_{\theta_0\theta_1}(1)$ on  the manifold $\calM$.

For two positive definite matrices $P$ and $Q$, let $P\succeq Q$ denote the L\"owner partial ordering.
That is, $P-Q$ is positive semi-definite: 
$P\succeq Q \Leftrightarrow P-Q\succeq 0$.

\subsection{Definition of the Fisher-Rao distance and Fisher-Rao geodesics} \Mylabel{sec:DefFR}

\begin{figure}%
\centering
\includegraphics[width=\columnwidth]{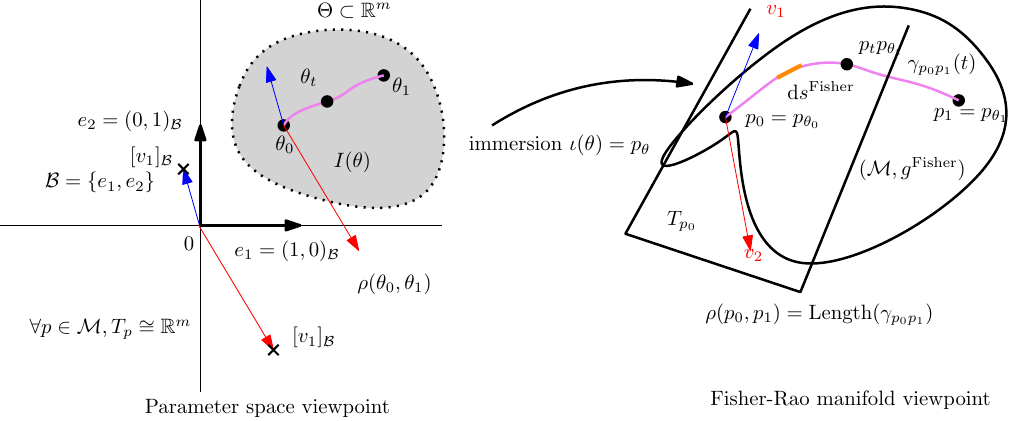}%

\caption{The parameter space $\Theta\subset\bbR^m$ of a $m$-dimensional statistical model $\calM=\{p_\theta\st\theta\in\Theta\}$ is considered as a Riemannian manifold using the immersion $\iota(\cdot)$ and the Fisher-Rao metric $g^\Fisher$.
Tangent spaces $T_p$ are vector spaces congruent to $\bbR^m$. }%
\Mylabel{fig:frdomain}%
\end{figure}

Let $(\calX,\calA,\nu)$ be a measure space~\cite{MeasureTheory-2004} with $\calX$ denoting the sample space, $\calA$ a $\sigma$-algebra and $\nu$ a positive measure.
Consider a {\em statistical model}  $\calM=\{P_\theta \st \theta\in\Theta\}$ of probability distributions $P_\theta$ dominated by $\nu$ 
with Radon-Nikodym densities $p_\theta=\frac{\dP_\theta}{\dnu}$, where $\Theta\in\bbR^m$ is $m$-dimensional open parameter space.
We assume that all statistical models are regular~\cite{calin2014geometric,IG-2016}: That is, there is a full rank immersion $\iota:\theta\rightarrow p_\theta$ such that
$\partial_1 p_\theta,\ldots, \partial_m p_\theta$ are linearly independent (required to define tangent planes when considering $\calM$ as a manifold).
See~\cite{Amari-1984} for the Finslerian differential geometry of irregular models.
 
Denote by $l_\theta(x)=\log p_\theta(x)$ the log-likelihood function and by $s_\theta(x)=\nabla_\theta l_\theta(x)$  the score function.
Under the Bartlett regularity conditions~\cite{Bartlett-1953-II,IG-2016} which allows to interchange integration with differentiation, we have the $m\times m$ {\em Fisher Information  Matrix} (FIM)
$I^\Fisher(\theta):=\Cov_\theta(s_\theta)=[I^\Fisher_{ij}(\theta)]$ defined equivalently as follows:
\begin{equation}
I^\Fisher_{ij}(\theta)=E_\theta[\partial_i l_\theta(x) \partial_j l_\theta(x)]=-E_\theta[\partial_i\partial_j l_\theta(x)].
\end{equation}
 
By considering the statistical model $\calM$ as a Riemannian manifold  $(\calM,g^\Fisher)$ equipped with the {\em Fisher information metric} $g^\Fisher$ expressed in the single global chart $\theta(\cdot)$ by the FIM, 
Rao~\cite{Rao-1945,Rao-1992} defined a distance $\rho_\calM(\theta_0,\theta_1)$ between two probability distributions 
$p_{\theta_0}$ and $p_{\theta_1}$ of a statistical model $\calM$ as the geodesic Riemannian distance.
When clear from context, we write for short $\rho(\theta_0,\theta_1)$.
We call Fisher connection $\bar\nabla$ the Levi-Civita metric connection induced by the Fisher metric $g$ with Christoffel symbols:
$$
\bar\Gamma_{ij}^k(\theta)=\frac{1}{2}\left(\partial_j g_{ik}(\theta)+\partial_i g_{jk}(\theta)-\partial_k g_{ij}(\theta)\right).
$$ 

The Fisher information matrix is invariant by reparameterization of the sample space but changes covariantly when reparameterizing with $\eta(\theta)$:
$$
I(\eta)= \Jac(\theta(\eta))^\top\,\times\, I(\theta(\eta))\,\times\, \Jac(\theta(\eta)),
$$ 
where $\Jac(\theta(\eta))=\left[\frac{\partial \theta_i}{\eta_j}\right]_{ij}$ is the Jacobian matrix of the transformation.

\begin{remark}
The Euclidean distance viewed as a Riemannian distance for the metric tensor $g^\Eucl$ is expressed in the Cartesian coordinates by the $d\times d$ identity matrix: $G(\theta)=I$. Thus under another reparameterization $\eta$, the Euclidean metric is expressed as $G(\eta)=J(\eta)^\top J(\eta)$ where $J(\eta)$ is a Jacobian matrix of transformation. 
\end{remark}

But the Fisher-Rao length element $\ds^\Fisher=\sqrt{\dtheta^\top\, I^\Fisher(\theta)\, \dtheta}$ and hence the Fisher-Rao distance are invariant under reparameterization of the parameter space.

Thanks to its geometric definition, the Fisher-Rao distance (or Rao distance~\cite{RaoEllipticalUB-Chen-2021}) is invariant under reparameterization of the parameter space.
That is, if we let $\eta=\eta(\theta)$ be a new parameterization of the statistical model $\calM$, we shall have
$$
\rho({\eta(\theta_0)},{\eta(\theta_1}))=\rho({\theta_0},{\theta_1}).
$$

See Figure~\ref{fig:frdomain} for an illustration of the geodesic distance on a Riemannian manifold with a single global chart.

\begin{example}
For example, consider three different parameterizations of the univariate normal distributions $N(\mu,\sigma)$ with probability density functions 
$p_{\mu,\sigma}(x)=\frac{1}{\sigma\sqrt{2\pi}}\, \exp(-\frac{1}{2}(\frac{x-\mu}{\sigma})^2)$:  $\lambda=(\mu,\sigma)$, $\lambda'=(\mu,\sigma^2)$, and $\lambda''=(\mu,\log\sigma)$. 
Then we have $\ds^\Fisher(\lambda,\dlambda)=\ds^\Fisher(\lambda',\dlambda')=\ds^\Fisher(\lambda'',\dlambda'')$, and the Rao geodesic distances do not depend on the chosen parameterization:
$$
\rho(({\mu_1,\sigma_1}),({\mu_2,\sigma_2}))=\rho(({\mu_1,\sigma_1^2}),({\mu_2,\sigma_2^2}))=\rho(({\mu_1,\log\sigma_1}),({\mu_2,\log\sigma_2})).
$$
However, the Fisher information matrices (FIMs) are different:
$$
I(\lambda) = \mattwotwo{\frac{1}{\sigma^2}}{0}{0}{\frac{2}{\sigma^2}},\quad
I(\lambda') = \mattwotwo{\frac{1}{\sigma^2}}{0}{0}{\frac{2}{\sigma^4}},\quad
I(\lambda'') = \mattwotwo{\frac{1}{\sigma^2}}{0}{0}{2}.
$$
Those Fisher information matrices $I(\lambda)$, $I(\lambda')$ and $I(\lambda'')$ define the same underlying Fisher Riemannian metric $g^\Fisher$ and length element.
The Fisher information matrix for bivariate normal distributions was reported in~\cite{sato1979geometrical}.
See Appendix~\ref{app:FIM} for symbolic computing of the Fisher information matrices.
\end{example}

Chentsov theorem~\cite{chentsov1982statiscal} and its generalization~\cite{dowty2018chentsov} states  that the Fisher metric is the unique Riemannian metric up to rescaling 
 that is invariant under sufficient statistics.
It follows that a statistical model $\calM=\{p_\theta\}$ admits a natural Riemannian geometric structure $(\calM,g^\Fisher)$.
 
Stigler~\cite{stigler2007epic} noticed that Hotelling historically first considered the Fisher-Rao distance in a handwritten note~\cite{Hotelling-1930} 
dated in 1929. Hotelling considered  the ``space of statistical parameters'' while Rao studied the ``population space.''
Interestingly, Mahalanobis mentioned the notion of ``statistical field'' in his celebrated paper introducing the Mahalanobis distance with tensor notations~\cite{chandra1936generalised}.
Nowadays, the term ``statistical manifold'' is used to either refer to a manifold of statistical models or to the pure dual geometric structure that was unravelled by Amari and Nagaoka~\cite{nagaoka1982differential,amari2000methods}.

Note that in general, Riemannian geodesics are not unique~\cite{godinho2012introduction}.
For example, geodesics on a sphere are great arc of circles, and two antipodal points on the sphere can be joined by infinitely  many half great circle geodesics. 
However, when the sectional curvatures of a Riemannian manifold $(M,g)$ are non-positive~\cite{bridson2013metric} (i.e., negative or zero), geodesics are guaranteed to be unique (see Hadamard spaces~\cite{bacak2014convex}).
Thus in general, Fisher geodesics may not be unique (an open question in~\cite{Eriksen-1987}) although we are not aware of a common statistical model used in practice exhibiting non-unique Fisher-Rao geodesics.

Let $(\calX,\Sigma_\calX)$ and $(\calY,\Sigma_\calY)$ be measurable spaces, and consider a stochastic kernel $K:\Sigma_\calY\times\calX\rightarrow [0,1]$~\cite{liese2008f}.
A statistical distance $D(P_1,P_2)$ is said monotone~\cite{liese2008f} if $D(KP_1,KP_2)\leq D(P_1,P_2)$.
The class of $f$-divergences~\cite{Csiszar-1967} are monotone~\cite{liese2008f}.

For example, consider the statistical model of categorical distributions on a sample space $\calX$ of size $d$.
Each categorical distribution is parameterized by a point  $p$ on the $(d-1)$-dimensional standard simplex $\Delta_{d-1}$ sitting in $\bbR^d$.
The Fisher-Rao distance between $p$ and $q$~\cite{miyamoto2023closed} is
\begin{equation}\label{eq:FRcat}
\rho(p,q)=2\,\arccos\left(\sum_{i=1}^d \sqrt{p_iq_i}\right).
\end{equation}
We may view categorical distributions $p=(p_1,\ldots, p_d)$ as normalized histograms with non-empty bins.
Consider reducing the dimension $d$ of categorical distributions $p$ by merging some bins to get categorical distributions $p'$ of $\Delta_{d'-1}$ with $d'<d$.
This merging bin operation can be interpreted as a deterministic Markov kernel. Then we have $\rho(p',q')\leq \rho(p,q)$.
In the extreme case of $d'=1$ (i.e., merging all bins to a single bin of value $\sum p_i=1$), we have $\rho(p',q')=\rho(1,1)=0\leq \rho(p,q)$.

\begin{property}\label{prop:frmonotone}
The Fisher-Rao distances are monotone.
\end{property}

\begin{proof}
We shall use the property of the Fisher information matrix.
Let $Y=t(X)$ be a statistic. Then $I_Y(\theta) \preceq I_X(\theta)$ with equality if and only if $Y$ is sufficient~\cite{keener2010theoretical}.
Thus we have the following inequality for corresponding Fisher metrics: $g_Y(\theta)\preceq g_X(\theta)$, and the corresponding Fisher-Rao length elements
 $\ds_Y$ and $\ds_X$ are such that
 $\ds_Y\leq \ds_X$ because $\ds_Y^2=\dtheta^\top g_Y(\theta) \dtheta \leq \ds_X^2=\dtheta^\top g_X(\theta) \dtheta$ since $g_X(\theta)-g_Y(\theta)\succeq 0$ (positive semi-definite).
Let $\gamma_X$ and $\gamma_Y$ be the geodesics with respect to Fisher metrics $g_X$ and $g_Y$, respectively.
We have
$$
\rho_Y(\theta_1,\theta_2)= \int_{\gamma_Y} \ds_Y \leq \int_{\gamma_X} \ds_Y 
\leq
\int_{\gamma_X} \ds_X=\rho_X(\theta_1,\theta_2). 
$$ 
Thus we have proved the monotonicity property of the Fisher-Rao distances: $\rho_Y(\theta_1,\theta_2)\leq \rho_X(\theta_1,\theta_2)$.
\end{proof}

\subsection{Computational tractability of Fisher-Rao distances}\Mylabel{sec:comptractability}
In order to calculate the Fisher-Rao distance $\rho_{\calM}$ for a given statistical model $\calM$, we need 
\begin{enumerate}
\item  to calculate the Fisher-Rao geodesics by solving the second-order ODE of Eq.~\ref{eq:RieOde} with boundary conditions, and 
\item  to integrate the Fisher length element $\ds^\Fisher$ along the Fisher-Rao geodesics $\gamma(t)$:
$$
\rho({\theta_0},{\theta_1})=\length^\Fisher(\gamma):=\int_0^1 \|\dot\gamma(t)\|_{\gamma(t)} \, \dt.
$$
\end{enumerate}
 
Although the Fisher-Rao distances are known for many statistical models~\cite{atkinson1981rao,reverter2003computing,MVRao-2005,miyamoto2023closed}, it is in general difficult to calculate in closed-form, even for the ubiquitous statistical model of multivariate normal distributions~\cite{MVRao-2005,Kobayashi-2023}.
Thus we need to derive approximation techniques and lower/upper bounds to use whenever the Fisher-Rao distances are not known in closed-form.

There is a particular case where $\nabla$-geodesics are easy to get in closed-form: When the Christoffel symbols  of the $\nabla$-connection vanish or equivalently when the Riemann-Christoffel curvature tensor is $0$ (i.e., $\nabla$ is a flat connection).
In that case, the geodesic ODE becomes:
\begin{equation}\Mylabel{eq:RieOdeFlat}
\forall k\in\{1,\ldots, m\},\quad \ddot\gamma_k(t)=0,
\end{equation}
and we can solve this ODE with boundary condition $\theta_0$ and $\theta_1$ as $\theta^k(t)=(1-t)\theta_0^k+t\theta_1^k$ for all $k\in\{1,\ldots, m\}$.
That is, $\nabla$-geodesics are straight line segments in the $\theta$-coordinate system.
We shall see that Hessian metrics $g$ are associated to a pair of canonical flat connections which prove useful to upper bound the Fisher-Rao distances (Proposition~\ref{prop:ubhessian}). 

The {\tt GeomStats} software package~\cite{miolane2020geomstats} for Riemannian geometry in machine learning provides a generic implementation of the  Fisher-Rao distance~\cite{le2023parametric} in the module {\tt information\_geometry} which calculates the Fisher information matrix (FIM) using  automatic differentiation  provided that the probability density function $p_\theta(x)$ is given explicitly.

The next section considers a generic upper bound on the Fisher-Rao distance that is obtained from 1D Fisher-Rao distances which can be calculated in structural closed forms.

\section{Generic closed-form cases and a canonical Fisher-Rao upper bound}\Mylabel{sec:generic}

\subsection{Fisher-Rao distances of uniparametric models}\Mylabel{sec:FR1D}

When the statistical model $\calM$ is uniparametric (i.e., order $m=1$), the Fisher metric is $g(\theta)=[g_{11}(\theta)]>0$, and the 
Fisher-Rao geodesic $\gamma_{p_{\theta_0}p_{\theta_1}(t)}$ between $p_{\theta_0}$ and $p_{\theta_1}$ is $\theta(s)$ (arclength parameterization) 
with $\dot\theta(s)=\frac{\dtheta}{\ds}(s)=1$. 
Hence, $\dtheta=\ds$ and we have
$$
\Length(\gamma)= \int_{\theta_0}^{\theta_1} \sqrt{g_{11}(\theta(t))}\, \dt= \int_{\theta_0}^{\theta_1} \sqrt{g_{11}(\theta)}\,\dtheta
= \left| h_g(\theta_0) - h_g(\theta_1) \right|,
$$ 
where $h_g(\theta)$ is an antiderivative of $\sqrt{g_{11}(\theta)}$, i.e., $h_g(\theta)=\int_{\theta_0}^\theta \sqrt{g_{11}(\theta)}\,\dtheta$ for any $\theta_0\in\Theta$.

\begin{proposition}[1D Fisher-Rao distance]\Mylabel{prop:fr1d}
The Fisher-Rao distance $\rho$ between two probability densities $p_{\theta_0}$ and $p_{\theta_1}$ of a uniparametric statistical model $\calM$ with Fisher-Rao metric $g^\Fisher=[g_{11}]$ is
$$
\rho(\theta_0,\theta_1)=\left| h_{g^\Fisher}(\theta_0) - h_{g^\Fisher}(\theta_1) \right|,
$$
where $h_g$ is an antiderivative of $\sqrt{g_{11}(\theta)}$.
\end{proposition}

In particular, we may consider uniparametric exponential families~\cite{barndorff2014information} with cumulant function $F(\theta)$ and Fisher information matrix $g^\Fisher(\theta)=[F''(\theta)]$:

\begin{proposition}[Fisher-Rao distance of uniparametric exponential families]\Mylabel{prop:FRef1d}
The Fisher-Rao between two densities of a uniparametric exponential family $\{p_\theta\}$ with cumulant function $F(\theta)$ is
\begin{equation}
\rho(\theta_0,\theta_1)=\left|h_F(\theta_1)-h_F(\theta_0)\right|,
\end{equation}
where $h_F(\theta)$ is an antiderivative of $\sqrt{F''(\theta)}$.
\end{proposition}

\begin{example}
For example, let us consider the exponential family~\cite{barndorff2014information}  of exponential distributions $\calM=\{p_\theta(x)=\theta \exp(-\theta x)\}$.
Let $t(x)=-x$ be the sufficient statistic, $\theta\in\Theta=\bbR_{>0}$, and $F(\theta)=-\log\theta$ be the cumulant function 
(strictly convex since $g_{11}(\theta)=F''(\theta)=\frac{1}{\theta^2}$).
We have $\sqrt{g_{11}(\theta)}=\sqrt{F''(\theta)}=\frac{1}{\theta}$ with antiderivative $h_F(\theta)=\log \theta$, and therefore we get the 
Fisher-Rao distance between two exponential distributions $p_{\theta_0}$ and $p_{\theta_1}$ as
$\rho_{\mathrm{Exponential}}(\theta_0,\theta_1)=\left|\log\frac{\theta_1}{\theta_2}\right|$.
\end{example}

In \S\ref{sec:frls}, we report the generic formula of the Fisher-Rao distances for the 1d elliptical or location-scale families.
Furthermore, we shall revisit the 1D Fisher-Rao distance case from the viewpoint of Hessian metrics in \S\ref{sec:HessianMetric}.

\subsection{Fisher-Rao distances for products of stochastically independent models}\Mylabel{sec:productmodel}

Let $(\calM_1,g_1),\ldots, (\calM_m,g_m)$ be $m$ Riemannian manifolds with corresponding length elements $\ds_1,\ldots,\ds_m$ and geodesic distances $\rho_{\calM_1},\ldots,\rho_{\calM_m}$.
Then the product manifold $\calM=\calM_1\times\ldots\calM_m$ is a Riemannian manifold with length element $\ds=\sqrt{\sum_i \ds_i^2}$, and  the Riemannian distance
 is $\rho_\calM(\theta,\theta')=\sqrt{\sum_{i=1}^m \rho_{\calM_i}^2(\theta^i,{\theta^i}')}$.

Thus let us consider statistical models which yield product manifolds:

\begin{proposition}[Fisher-Rao distance of products of stochastically independent models]\Mylabel{prop:FRindepsto}
For $m$ independent statistical models $\calM_1=\{p_{\theta^1}(x_1)\}$ , $\ldots$, $\calM_m=\{p_{\theta^m}(x_m)\}$ with corresponding 
 variables $x_i\in\bbR^{d_i}$ and parameters $\theta^i\in\bbR^{m_i}$ (with $d=\sum_i d_i$ and $m=\sum_i m_i$), 
and Fisher-Rao's distances $\rho_{\calM_1}$, $\ldots$, $\rho_{\calM_m}$, 
we have the Fisher-Rao distance on the product  model 
$\calM=\{p_\theta(x)\propto \prod_{i=1}^m p_{\theta^i}(x_i) \st \theta\in\bbR^m\}$ which is given by 
\begin{equation}
\rho_{\calM}(\theta_0,\theta_1)=\sqrt{\sum_{i=1}^m \rho_{\calM_i}^2(\theta_0^i,\theta_1^i)}.
\end{equation}
\end{proposition}

This technique allows for example to derive the Fisher-Rao distance between normal distributions with diagonal covariance matrices~\cite{krzanowski1996rao}.
Indeed, when the covariance matrices of normal distributions are diagonal, we have $\Cov(x_i,x_j)=0$ when $i\not=j$, and the $d$-variate normal distributions with diagonal covariance matrices can be interpreted as a product of $d$ univariate normal distributions:
$$
p_{\mu,\Sigma=\diag(\sigma_{11}^2,\ldots,\sigma_{dd}^2)}= \frac{1}{(2\pi)^{\frac{d}{2}}\, \sqrt{|\Sigma|}}\, \exp\left(-\frac{1}{2}(x-\mu)^\top\, \Sigma^{-1} (x-\mu)\right)
=
 \prod_{i=1}^d p_{\mu_i,\sigma_{ii}^2}(x_i),
$$
where  $p_{\mu,\sigma^2}(x)=\frac{1}{\sqrt{2\pi\sigma^2}}\exp(-\frac{1}{2} \frac{(x-\mu)^2)}{\sigma^2})$.

\subsection{The Fisher-Manhattan upper bound for the Fisher-Rao distances}\Mylabel{sec:FisherManhattan}

\begin{figure}%
\centering
\includegraphics[width=0.75\columnwidth]{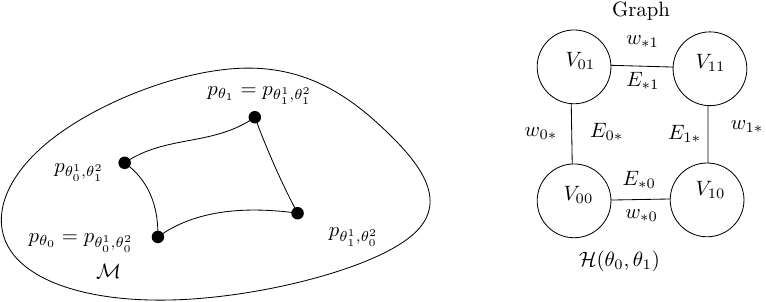}%
\caption{Illustration of the Fisher-Manhattan upper bound for a bidimensional statistical model ($m=2$): We calculate the 1D Fisher-Rao distances along the edges of the square and then find the shortest path on the weighted graph between node $\theta_0$ and node $\theta_1$.}%
\Mylabel{fig:hypercube}%
\end{figure}

To get an upper bound on the Fisher-Rao distance of $m$-order statistical models $\calM$, we shall consider the closed-form formula for 1D Fisher-Rao distances of its  uniparametric submodels (obtained by fixing all but a single parameter coordinate) and then use the triangle inequality property of the Fisher-Rao metric distances.

Given two densities $p_{\theta_0}$ and $p_{\theta_1}$ of $\calM$ with $\theta_0=(\theta_0^1,\ldots,\theta_0^m)$ and 
$\theta_1=(\theta_1^1,\ldots,\theta_1^m)$,  consider a $m$-dimensional hypercube graph $\calH(\theta_0,\theta_1)$ with the following $2^m$ vertices: 
$$
\theta_{a_1,\ldots,a_m}=(\theta_{a_1,\ldots,a_m}^1,\ldots, \theta_{a_1,\ldots,a_m}^m),
$$
with
$$
\theta_{a_1,\ldots,a_m}^i=(\theta_0^i)^{a_i}\, (\theta_1^i)^{1-a_i},
$$
for $(a_1,\ldots,a_m)\in \{0,1\}\times\ldots\times\{0,1\}$.
We have $\theta_{0,\ldots,0}=\theta_0$ and $\theta_{1,\ldots,1}=\theta_1$:
That is, if $\theta_0$ is one  corner of the hypercube $\calH(\theta_0,\theta_1)$ then $\theta_1$ is the opposite corner of the hypercube, and vice-versa.
See Figure~\ref{fig:hypercube} for an illustration in 2D.
An edge $e=((a_1,\ldots,a_m),(a_1',\ldots,a_m'))\in\calH$ of the hypercube  has its two vertex notations which differ exactly in only one position $a_i\not=a_i'$.
That is, the nodes of the hypercube are labeled using the Hamming code~\cite{nielsen2016introduction} so that for an edge $e$ there exists a single index $i$ such that $a_i'=1-a_i$ and $a_j'=a_j$ for all other indices $j\not=i$. Thus we can write the edge $e$ as $e_{a_1\ldots a_{i-1}*a_{i+1}\ldots a_m}$ where the star symbol indicates the $i$'th axis of the edge.

We consider the 1D statistical models $\calM_e$'s where only the $i$-th parameter of $\calM$ is allowed to vary, and calculate its Fisher information 
$g_{11}^e(\theta^i)$ and get the Fisher-Rao distance using Proposition~\ref{prop:FRef1d}.
Overall, we compute the 1D Fisher-Rao distances along the $m2^{m-1}$ edges  of the hypercube $\calH(\theta_0,\theta_1)$.
Then we compute the shortest distance $\mathrm{ShortestPath}_{\calH(\theta_0,\theta_1)}(\theta_0,\theta_1)$ between vertex $\theta_{0,\ldots,0}=\theta_0$ and vertex  $\theta_{1,\ldots,1}=\theta_1$ on the edge-weighted hypercube graph where edges have all non-negative weights using Dijkstra's algorithm which runs in $O(m2^m)$ time.

Using the triangle inequality property of the Fisher-Rao distances, we get the upper bound $\rho(\theta_0,\theta_1)\leq \mathrm{ShortestPath}_{\calH(\theta_0,\theta_1)}(\theta_0,\theta_1)$.
We call this class of upper bounds the generic Fisher-Manhattan upper bound.

\begin{proposition}[Fisher-Manhattan upper bound]\Mylabel{prop:FisherManhattan}
The Fisher-Manhattan upper bound on a $m$-dimensional statistical model can be calculated in $O(m2^m)$ time and is tight when the statistical model is a stochastically independent product model.
\end{proposition} 

We can consider the 1D Fisher-Rao geodesics corresponding to the edges of the hypercube as 1D submanifolds of the Fisher-Rao manifold $\calM$.
Those  1D Fisher-Rao geodesics can  either be totally geodesic submanifolds or not (see Section~\ref{sec:totallygeo}).
When all 1D Fisher-Rao geodesics are totally geodesic submanifolds, the Fisher-Manhattan upper bounds are tight.

Figure~\ref{fig:FisherManhattanNormal} illustrates the Fisher-Manhattan bound for the Fisher-Rao manifold of univariate normal distributions $\calM=\{N(\mu,\sigma) \st \mu\in\bbR, \sigma\in\bbR_{>0}\}$ ($m=2$).
In that case, the exact Fisher-Rao distance amounts to a Poincar\'e hyperbolic distance after reparameterization~\cite{UniElliptical-Mitchell-1988}.

\begin{figure}%
\centering
\includegraphics[width=0.75\columnwidth]{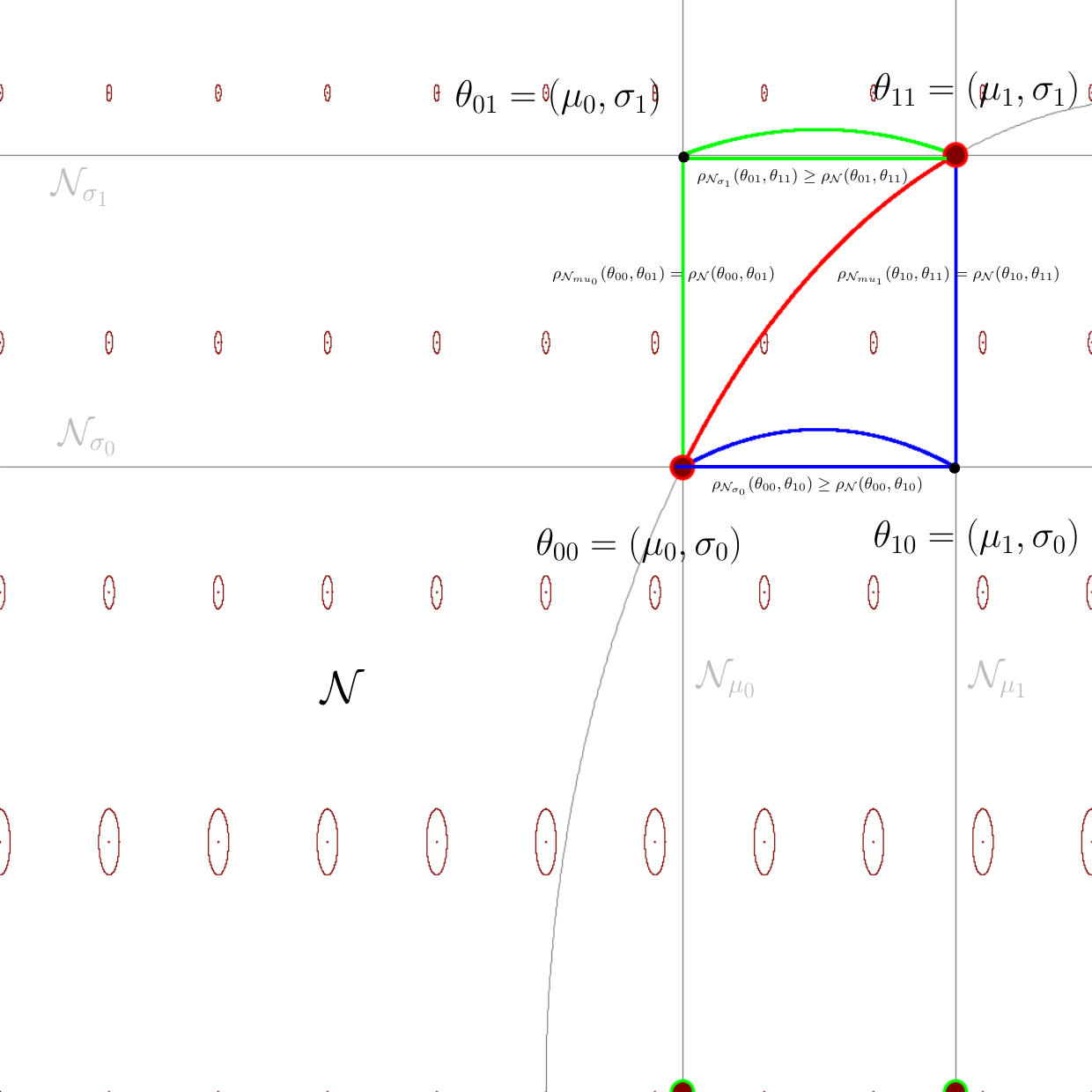}%

\caption{Illustrating the Fisher-Manhattan upper bound for the Fisher-Rao manifold of univariate normal distributions $\mathcal{N}$ ($m=2$):
The Fisher-Rao geodesic between $N(\mu_0,\sigma_0)$ and $N(\mu_1,\sigma_1)$ is shown in red.
The Fisher-Manhattan upper bound is derived from the straight-edge square $((\mu_0,\sigma_0),(\mu_0,\sigma_1),(\mu_1,\sigma_1),(\mu_1,\sigma_0))$.
The Fisher-Manhattan upper bound is $\min(\rho((\mu_0,\sigma_0),(\mu_0,\sigma_1))+\rho((\mu_0,\sigma_1),(\mu_1,\sigma_1)),
(\mu_0,\sigma_0),(\mu_1,\sigma_0)+\rho((\mu_1,\sigma_0),(\mu_1,\sigma_1)))$, i.e., the minimum length of the two green and blue straight-edge path lengths. 
Tissot indicatrices of the Fisher metric are displayed at 2D grid locations.
The submanifolds $\mathcal{N}_{\sigma_0}$ and $\mathcal{N}_{\sigma_1}$ are not totally geodesics and the 
submanifolds $\mathcal{N}_{\mu_0}$ and $\mathcal{N}_{\mu_1}$ are totally geodesics.
}%
\Mylabel{fig:FisherManhattanNormal}%
\end{figure}

\section{Generic approximation schemes for the Fisher-Rao distances}\Mylabel{sec:approxscheme}

\subsection{Approximation of the Fisher-Rao lengths of curves}\Mylabel{sec:approxFRlength}

When the Fisher-Rao geodesics are not available in closed form, we may measure the Fisher-Rao length of any smooth curve $c(t)$ with explicit parameterization:
This yields an upper bound on the Fisher-Rao geodesic distance. 
For example, we may consider the $\theta$-straight curve $c(t)=\mathrm{LERP}(p_{\theta_0},p_{\theta_1};t)$ obtained by linear interpolation of its parameter endpoints: 
$\theta(t)=\theta_0 + t (\theta_1-\theta_0)$.
Since the Fisher-Rao distance is defined as the infimum of the Fisher-Rao lengths of curves connecting $p_{\theta_0}$ and $p_{\theta_1}$ according to Eq.~\ref{eq:inflength}, we have
$$
\rho(\theta_0,\theta_1)\leq \Length(c(t)).
$$
Thus the Fisher-Rao distance approximation problem can be reduced to calculate or approximate the length of an arbitrary curve $c(t)$ for $t\in [0,1]$ which is closed to the Fisher-Rao geodesics.

In general, we may sample the curve $c(t)$ at $T\geq 2$ positions $t_0=0,t_1,\ldots, t_{T-2}, t_{T-1}=1$, and approximate the Fisher-Rao length of $c$ as
$$
\widetilde{\Length}(c) = \sum_{i=0}^{T-2} \rho(\theta_{t_i},\theta_{t_{i+1}}).
$$

Whenever $t_{i+1}-t_i$ is small enough (e.g., let $t_i=\frac{i}{T-1}$ so that $t_{i+1}-t_i=\frac{1}{T-1}$), 
we may use the property that the Fisher-Rao lengths $\rho(\theta_{t_i},\theta_{t_{i+1}})$  can be well approximated from any $f$-divergence~\cite{fdiv-AliSilvey-1966,Csiszar-1967},
where a $f$-divergence between two  probability densities $p(x)$ and $q(x)$ is defined for a convex generator $f(u)$ strictly convex at $u=1$ as
$$
I_f(p:q)=\int p(x)\, f\left(\frac{q(x)}{p(x)}\right)\dmu(x).
$$
When $f(u)=-\log u=f_\KL(u)$, we recover the Kullback-Leibler divergence, i.e., the relative entropy between $p(x)$ and $q(x)$: 
$$
D_\KL(p:q)=I_{f_\KL}(p:q)=\int p(x)\, \log\frac{p(x)}{q(x)}\,\dmu(x).
$$

For an arbitrary smooth $f$-divergence\footnote{This does not include the total variation distance which is the only metric $f$-divergence obtained for the generator $f_\TV(u)=|u-1|$.}, we have~\cite{calin2014geometric}:
\begin{equation}
I_f(p_{\theta_0}:p_{\theta_1}) = \frac{f''(1)}{2}\rho^2(\theta_0,\theta_1) + o(\rho^2(\theta_0,\theta_1)).
\end{equation}

Thus when $|\theta-\theta'|$ is small enough, we approximate $\rho(\theta,\theta')$ by
\begin{equation}
\rho(\theta,\theta')\approx\tilde \rho(\theta,\theta')= \sqrt{\frac{2}{f''(1)}\, I_f(p_{\theta}:p_{\theta'})}.
\end{equation}

\begin{figure}
\centering
\begin{tabular}{ccc} 
\includegraphics[width=0.3\textwidth]{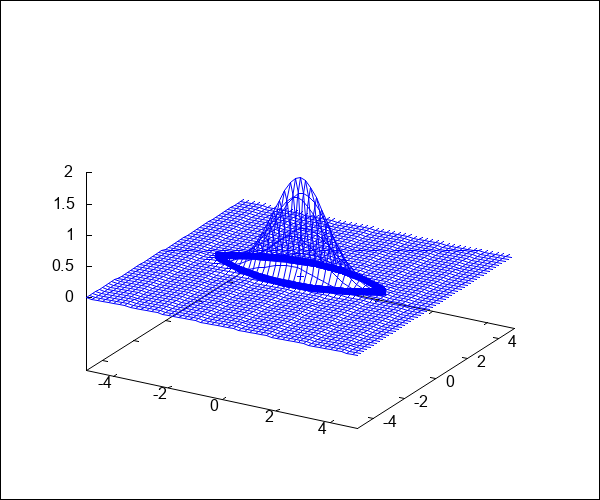} &
\includegraphics[width=0.3\textwidth]{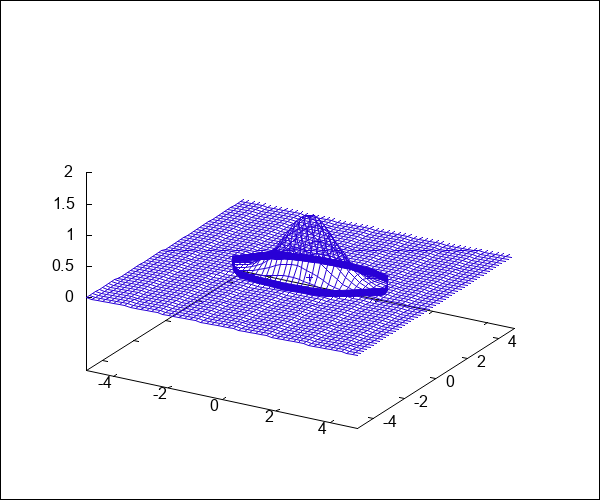} &
\includegraphics[width=0.3\textwidth]{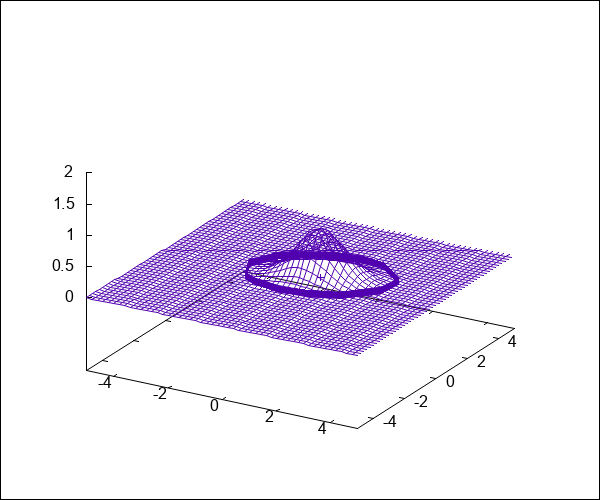} \\
\includegraphics[width=0.3\textwidth]{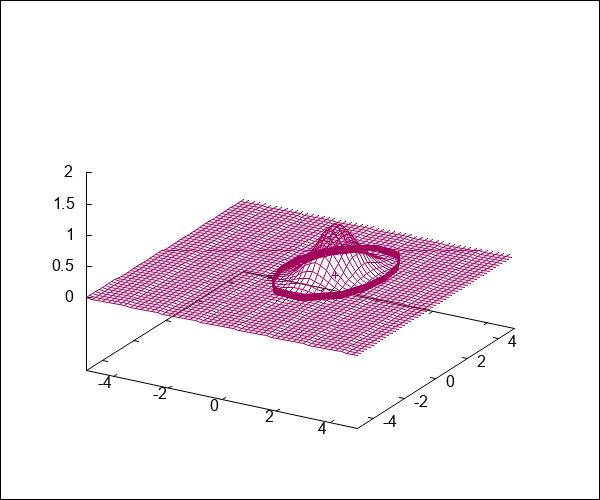} &
\includegraphics[width=0.3\textwidth]{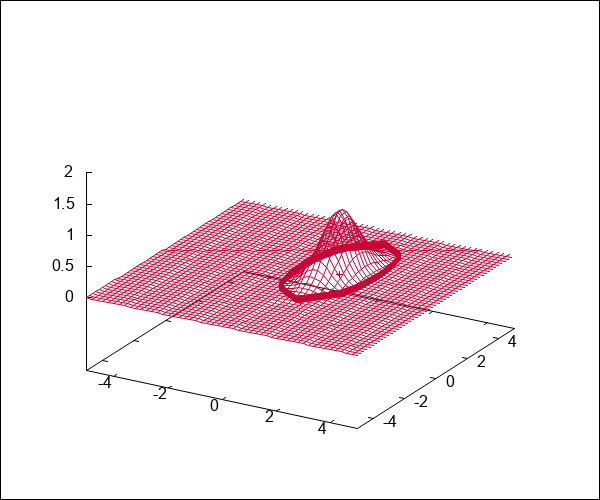} &
\includegraphics[width=0.3\textwidth]{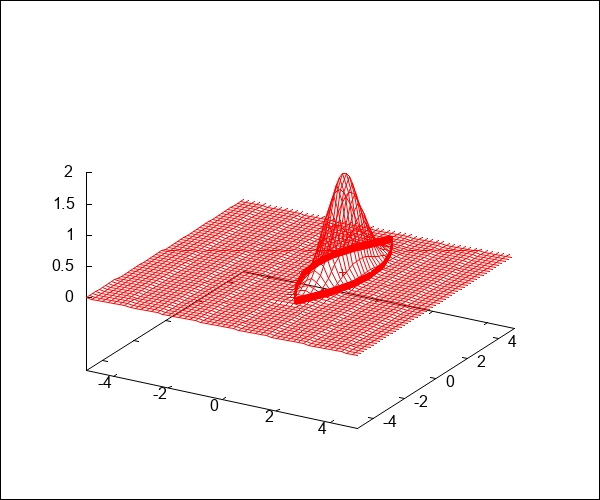} 
\end{tabular}

\caption{Fisher-Rao geodesic interpolation between two bivariate normal distributions with respective densities shown in top left (blue) and bottom right (red).}
\Mylabel{fig:bivarnormalinterpolation}

\end{figure}

\begin{proposition}[Approximation of the Fisher-Rao length of a curve]\Mylabel{prop:approxFRlength}
 The Fisher-Rao length of a smooth curve  $c: [0,1]\rightarrow\calM$ of a Fisher-Rao manifold $\calM=\{p_\theta\}$ can be approximated by
$$
\widetilde{\Length}(c)=  \sqrt{\frac{2}{f''(1)}} \left( \sum_{i=0}^{T-2} \sqrt{ I_f( p_{\theta_{t_i}} :p_{\theta_{t_{i+1}}} ) } \right),
$$
where $t_0=0,\ldots, t_{T-1}=1$ are $T$ chosen steps in $[0,1]$ such that  all $t_{i+1}-t_i$'s are small enough.
\end{proposition}

When choosing the Jeffreys divergence $D_J(p,q)=I_f(p:q)$, a  $f$-divergence with $f(u)=(u-1)\log u=f_J(u)$ and $f''_J(1)=2$, we get the following corollary:
\begin{corollary}
The Fisher-Rao length of a smooth curve  $c: [0,1]\rightarrow\calM$ of a Fisher-Rao manifold $\calM=\{p_\theta\}$ can be approximated by
\begin{equation}
\widetilde{\Length}(c)=  \left( \sum_{i=0}^{T-2} \sqrt{D_J(p_{\theta_{t_i}}:p_{\theta_{t_{i+1}}})} \right).
\end{equation}
\end{corollary}

\begin{example}
In practice, we may have some $f$-divergences between two distributions in closed-form although the Fisher-Rao distance is not known in closed-form.
For example, the Jeffreys divergence (i.e., the symmetrized Kullback-Leibler divergence) between multivariate normal distributions is available in closed form but not the corresponding Fisher-Rao distance~\cite{Kobayashi-2023}.
Figure~\ref{fig:bivarnormalinterpolation} displays the Fisher-Rao geodesic interpolation between two multivariate normal distributions (MVNs) 
$N_1=N\left(\vectortwo{0}{0},\mattwotwo{1}{0}{0}{0.1}\right)$ and $N_2=N\left(\vectortwo{1}{1},\mattwotwo{0.1}{0}{0}{1}\right)$ obtained using the method described in~\cite{Kobayashi-2023}. 
Since computing the definite integral of the Fisher-Rao distance is not known in closed-form for MVNs, we may approximate its length by the above sampling method with $c(t)=\gamma(t)$.
\end{example}

When $p_{\theta_0}$ is close to $p_{\theta_1}$ (i.e., $\theta_1=\theta_0+\Delta\theta$ with $\|\Delta\theta\|$ small), 
we may further approximate $I_f(p_{\theta_0}:p_{\theta_1}$ by $\frac{f''(1)}{2} \widetilde{\ds}^2_\Fisher(\theta_0,\Delta\theta)$.

For example, the Fisher-Rao length element of the Fisher-Rao multivariate normal  manifold is 
$$
\ds_\Fisher^2((\mu,\Sigma),(\dmu,\dSigma))= \dmu^\top \Sigma^{-1} \dmu + \frac{1}{2}\tr\left(\left(\Sigma^{-1}\dSigma\right)^2\right),
$$ 
and we approximate it for $\Delta\theta=(\mu_1-\mu_0,\Sigma_1-\Sigma_0)$ by
$$
\widetilde{\ds}_\Fisher^2((\mu,\Sigma),(\mu_1-\mu_0,\Sigma_1-\Sigma_0))=
(\mu_1-\mu_0)^\top\Sigma^{-1}_0 (\mu_1-\mu_0) +\frac{1}{2}\tr\left(\left(\Sigma^{-1}_0 (\Sigma_1-\Sigma_0))\right)^2\right).
$$

\subsection{Approximations when the Fisher-Rao  (pre)geodesics are available}\Mylabel{sec:metricspace}

We assume the Fisher-Rao manifold $(\calM,g^\Fisher)$ to be a connected and smooth Riemannian manifold so that
by the Hopf-Rinow theorem~\cite{godinho2012introduction}, $(\calM,\rho)$ is a complete metric space.
In that case, there exists a length minimizing geodesic $\gamma_{\theta_1\theta_0}(t)$ connecting any two points $p_{\theta_0}=\gamma_{\theta_0\theta_1}(0)$ and  $p_{\theta_1}=\gamma_{\theta_0\theta_1}(1)$ on  the manifold, and we have $\rho(\theta_0,\theta_1)=\Length(\gamma_{\theta_0\theta_1})$.

The Fisher-Rao Riemannian distance $\rho(\cdot,\cdot)$ is thus a metric distance that satisfies the following identity:
\begin{equation}
\forall s\in [0,1],\forall t\in [0,1],\quad \rho(p_s,p_t)=|s-t|\, \rho(p_0,p_1),
\end{equation}
where $p_t=\gamma_{p_0p_1}(t)$ denotes the geodesic arclength parameterization passing through the geodesic endpoints $p_0$ and $p_1$ (constant speed geodesic).
Hence, we have 
$$
\forall t\in [0,1],\quad \rho(p_0,p_t)=t \, \rho(p_0,p_1),
$$
and therefore, we get the following identity:
\begin{equation}\Mylabel{eq:scalerho}
\forall t\in (0,1],\quad \rho(p_0,p_1)=\frac{1}{t}\,\rho(p_0,p_t),
\end{equation}
where $p_t=\gamma_{\theta_0\theta_1}(t)$.
Notice that in practice, we have to take into account the numerical precision of calculations.

\begin{example}
For example, on the Euclidean manifold, we have $\rho(p_s,p_t)=\|p_s-p_t\|_2=|s-t|\, \|p_0-p_1\|_2$,
and $p_t=p_0+t(p_1-p_0)$.
Thus, we have $\rho(p_0,p_t)=t\, \|p_1-p_0\|_2$ and $\rho(p_0,p_1)=\frac{1}{t}\,\rho(p_0,p_t)=\|p_1-p_0\|_2$.
\end{example}

Thus when the Fisher-Rao geodesics $p_{\theta_t}=\gamma_{p_{\theta_0}p{\theta_1}}(t)$ 
passing through two points $p_{\theta_0}=\gamma_{p_{\theta_0}p_{\theta_1}}(0)$ and $p_{\theta_1}=\gamma_{p_{\theta_0}p_{\theta_1}}(1)$  are known explicitly (i.e., Riemannian geodesics with boundary conditions), we have (Figure~\ref{fig:approxFR}):
$$
\rho(\theta_0,\theta_1)=\frac{1}{|s-t| } \, \rho(\theta_s,\theta_t),\quad \forall (s,t)\in [0,1]^2, s\not=t. 
$$

In particular, for $s=0$ and small values of $t>0$ we get:
\begin{equation}\Mylabel{eq:approxrhofdiv}
\rho({\theta_0},{\theta_1})=\frac{1}{t}\,\rho({\theta_0},\theta_t) \approx \frac{1}{t}\, \sqrt{\frac{2}{f''(1)}\, I_f(p_{\theta_0}:\gamma_{p_{\theta_0}p_{\theta_1}}(t))}.
\end{equation}

\begin{figure}%
\centering
\includegraphics[width=0.5\columnwidth]{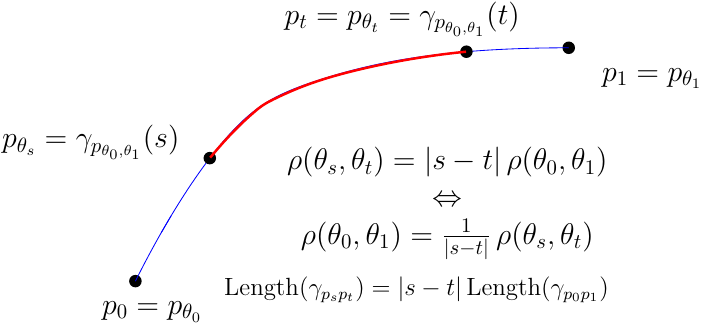}%
\caption{Illustration of the metric property of Fisher-Rao metric distances.}%
\Mylabel{fig:approxFR}%
\end{figure}

We may further approximate $\rho({\theta_0},\theta_t)$ by the Fisher length element at $\theta_0$ with direction $\Delta\theta(t)=\theta_t-\theta_0$:
$$
\rho({\theta_0},\theta_t) \approx \ds(\theta_0,\Delta\theta(t)).
$$

Thus we get the following proposition:

\begin{proposition}[Fisher-Rao distance approximation]\Mylabel{prop:frepsgeo}
When the Fisher-Rao geodesic $\gamma_{p_{\theta_0}p_{\theta_1}}(t)$ with boundary values is known in closed-form, 
we can approximate the Fisher-Rao distance 
$\rho(\theta_0,\theta_1)$ by
$$
\tilde\rho_\eps(\theta_0,\theta_1) = \frac{1}{\eps}\, \widetilde{\ds}(\theta_0,\Delta\theta(\eps)), 
$$ 
where $\widetilde{\ds}(\theta_0,\Delta\theta(\eps))^2=(\theta_\eps-\theta_0)^\top\, I^\Fisher(\theta_0)\, (\theta_\eps-\theta_0)$ is the finite difference (not infinitesimal) length element approximating the infinitesimal length element with $\theta_\eps=\theta(\gamma_{p_{\theta_0}p_{\theta_1}}(\eps))$, and
 $\eps>0$ an arbitrary small value.
\end{proposition}

Alternatively, we may consider the reference parameter to be $\theta_s$ for $s\in [0,1-\eps)$ so that
$$
\tilde\rho_{\theta_s,\eps}(\theta_0,\theta_1) = \frac{1}{\eps}\, \widetilde{\ds}(\theta_s,\Delta_s\theta(\eps)),
$$
with $\ds(\theta_s,\Delta\theta(\eps))^2=(\theta_{s+\eps}-\theta_s)^\top\, I^\Fisher(\theta_s)\, (\theta_{s+\eps}-\theta_0)$.
Notice that  the $\epsilon$ is not related to a guaranteed $(1+\epsilon)$-approximation of the Fisher-Rao distance~\cite{FisherRaoMVN-Nielsen-2023}, but is rather a tuning parameter that may yield an approximation $\tilde\rho_\eps$ which can be larger or smaller than the true Fisher-Rao distance.
Thus we may amortize the error by averaging at $k$ positions $\theta_{s_i}$: 
$$
\tilde\rho_{k,\eps}(\theta_0,\theta_1) = \frac{1}{k}\sum_{i=1}^k  \tilde\rho_{\theta_{s_i},\eps}(\theta_0,\theta_1) .
$$

Let us remark that even when the Fisher-Rao distance is known in closed-form, the above approximation may be computationally useful since it may be calculated faster than the exact closed-form formula as illustrated by the following example:

\begin{example}
Consider the Fisher-Rao metric of centered $d$-variate normal distributions $\calN_\mu=\{p_{\mu,\Sigma}\st \Sigma\succ 0\}$ (parameter space is the symmetric positive-definite matrix cone). The Fisher information metric is  a scaled trace metric, and the infinitesimal Fisher length element is $\ds^2_\Sigma(\Sigma,\dSigma)=\frac{1}{2}\tr((\Sigma^{-1}\dSigma)^2)$.
The Fisher-Rao geodesic $\gamma_{\Sigma_0\Sigma_1}(t)$ is given by:
\begin{equation}\Mylabel{eq:FisherRaoSPDgeo}
\Sigma(t)=\Sigma(\gamma_{\Sigma_0\Sigma_1}(t))=\Sigma_0^{\frac{1}{2}}(\Sigma_0^{-\frac{1}{2}}\Sigma_1\Sigma_0^{-\frac{1}{2}})^t \Sigma_0^{\frac{1}{2}},
\end{equation}
and the Riemannian distance is known in closed-form~\cite{Siegel-1964,James-1973}:
\begin{equation}\Mylabel{eq:FisherRaoSPD}
\rho(\Sigma_0,\Sigma_1) = \sqrt{\frac{1}{2} \sum_{i=1}^d \log^2 \lambda_i(\Sigma_0^{-1}\Sigma_1)},
\end{equation}
where the $\lambda_i(X)$'s denote the $i$'th smallest eigenvalue of a matrix $X$.

Applying Proposition~\ref{prop:frepsgeo}, we have $\widetilde{\ds}(\Sigma_0,\Sigma_\eps-\Sigma_0)=\frac{1}{\eps \sqrt{2}} \sqrt{\tr\left(\Sigma_0^{-1}(\Sigma_0-\Sigma_\eps)\right)^2}$ and get
\begin{equation}
\rho(\Sigma_0,\Sigma_1)\approx \rho_\eps(\Sigma_0,\Sigma_1)=\frac{1}{\eps \sqrt{2}} \sqrt{\tr\left(\Sigma_0^{-1}(\Sigma_0-\Sigma_\eps)\right)^2}.
\end{equation}

Notice that the Jeffreys divergence between the centered $d$-variate Gaussians $p_{\mu,\Sigma_0}$ and $p_{\mu,\Sigma_1}$   is
$$
D_J(p{\mu,\Sigma_0},p_{\mu,\Sigma_1})=\frac{1}{2}\tr(\Sigma_1^{-1}\Sigma_0+\Sigma_0^{-1}\Sigma_1)-d,
$$ 
and we have $\widetilde{\ds}(\Sigma_0,\Sigma_\eps)\approx \sqrt{D_J(p_{\mu,\Sigma_0},p_{\mu,\Sigma_\eps})}$.

For example, consider $\Sigma_0 =  \mattwotwo{3/2}{1}{1}{1}$ and 
 $\Sigma_1 = \mattwotwo{2}{1}{1}{1}$, then, 
$\Sigma_1 \Sigma_0^{-1} =  \mattwotwo{2}{-1}{0}{1}$ (not a positive-definite matrix!), and 
$\lambda_1(\Sigma_1 \Sigma_0^{-1})=1$, $\lambda_2(\Sigma_1 \Sigma_0^{-1})=2$.
Therefore, we have $\rho(\Sigma_0,\Sigma_1)=\frac{\log 2}{\sqrt{2}}$.

Although we have a closed-form formula for the Rao distance given in Eq.~\ref{eq:FisherRaoSPD}, the approximation scheme only requires to compute the interpolant on the geodesic given in Eq.~\ref{eq:FisherRaoSPDgeo}. This geodesic interpolant $\Sigma_t$ can be calculated from an eigendecomposition or approximated numerically using the matrix arithmetic-harmonic mean described in~\cite{nakamura2001algorithms} (AHM) which converges fast in quadratic time.
Thus we may bypass the eigendecomposition and get an approximation $\tilde\Sigma_\eps$ of $\Sigma_\eps$ from which we can calculate 
the Fisher-Rao distance approximation using Proposition~\ref{prop:frepsgeo}.
The advantage is to get Fisher-Rao distance approximations in faster time than applying the closed-form formula of Eq.~\ref{eq:FisherRaoSPD}.
Theoretically, the complexity of a SVD is of the same order of the complexity of a matrix inversion which is of the same order of the matrix multiplication~\cite{demmel2007fast}: $O(n^\omega)$ with $\omega=2.371866$~\cite{duan2023faster}.
\end{example}

Next, let us consider the approximation of the Fisher-Rao distance between multivariate normal distributions

\begin{example}
Let $\calN=\{p_{\mu,\Sigma}(x) \st (\mu,\Sigma)\in \bbR^d\times \bbP(\bbR,d)\}$ be the family of $d$-variate normal distributions.
The squared Fisher-Rao line element induced   the multivariate normal family~\cite{Skovgaard-1984} is
\begin{equation}
\ds^2(\mu,\Sigma) = \dmu^\top \Sigma^{-1} \dmu + \frac{1}{2}\tr\left(\left(\Sigma^{-1}\dSigma\right)^2\right).
\end{equation}
Therefore, we have
$$
\rho((\mu_0,\Sigma_0),(\mu_0,\Sigma_1))\approx \rho_\eps(\Sigma_0,\Sigma_1)=\frac{1}{\eps}\, \widetilde{\ds}((\mu_0,\Sigma_0),(\mu_\eps,\Sigma_\eps)),
$$
where $(\mu_\eps,\Sigma_\eps)$ is calculated using the Riemannian submersion method of Kobayashi~\cite{Kobayashi-2023}.
\end{example}

\begin{remark}
Notice that the length element of the Fisher information metric of the family of $d$-variate Wishart distributions with scale matrix $V$ is proportional to the trace metric~\cite{James-1973,EWishart-2023}:
$$
\ds_\Wishart^2 = \frac{d}{2} \tr\left(\left(V^{-1}\dm\right)^2\right).
$$
In general, if $\rho_g$ denotes the geodesic distance with respect to metric $g$, then $\sqrt{s}\rho_g$ is the geodesic distance with respect to scaled metric $s g$ for any $s>0$. Thus the Wishart Fisher-Rao distance is $\rho_{\mathrm{Wishart}}(V_1,V_2)=\sqrt{\frac{d}{2}}\, \rho_{\mathrm{SPD}}(V_1,V_2)$
 or $\rho_{\mathrm{Wishart}}(V_1,V_2)=\sqrt{d}\, \rho_{\mathcal{N}_0}(V_1,V_2)$ where $\mathcal{N}_0$ is the Fisher-Rao manifold of normal distributions centered at the origin.
\end{remark}

Thus the main limitation of this technique is to calculate $\theta_\eps$ (or a good approximation) for small enough values of $\eps>0$.


\subsection{Guaranteed approximations of Fisher-Rao distances with arbitrary relative errors}\Mylabel{sec:metric}

\begin{algorithm}
\KwIn{Probability densities $p_{\theta_0}$ and $p_{\theta_1}$ on $\calM=\{p_\theta\}$}
\KwIn{Closed-form  geodesic $\gamma_{p_{\theta_0}p_{\theta_1}}(t)$ with arclength parameterization}
\KwIn{Maximum multiplicative error $\eps>0$ }
\KwOut{A guaranteed approximation $\tilde\rho_\eps(\theta_0,\theta_1)\leq (1+\eps)\rho(\theta_0,\theta_1)$}
\tcc{Get lower bound $\rho_L$ and upper bound $\rho_U$ on the Fisher-Rao distance}
$\rho_L=\mathrm{LowerBound}(\theta_0,\theta_1)$\;
$\rho_U=\mathrm{UpperBound}(\theta_0,\theta_1)$\;
\If{$\frac{\rho_U}{\rho_L}>1+\eps$}{
\tcc{Recursion using the geodesic midpoint}
\Return $2\times \mathrm{ApproxFisherRao}({\theta_0},\frac{1}{2}(\theta_1+\theta_0),\eps)$\; }\Else{\tcc{Terminal case of recursion}
\Return $\rho_U$}
\caption{Recursive algorithm $\mathrm{ApproxFisherRao}({\theta_0},\theta_1,\eps)$ for getting a $(1+\eps)$-approximation of the Fisher-Rao distance 
when the geodesics are available in closed-form.}
\Mylabel{algo:multerrorgeo}
\end{algorithm}

When the Fisher-Rao geodesic $\gamma(t)$ with boundary conditions $p_{\theta_0}$ and $p_{\theta_1}$ is available in closed form,
and that both tight lower $\rho_L$ and upper $\rho_U$ bounds are available, we may guarantee a $(1+\eps)$-approximation of the Fisher-Rao distance for any $\eps>0$ as described by the method 
$\mathrm{ApproxFisherRao}({\theta_0},\theta_1,\eps)$ in 
Algorithm~\ref{algo:multerrorgeo}.

We may relax the requirement of closed-form Fisher-Rao geodesics to Fisher-Rao pregeodesics which do not require arc length parameterization.
A pregeodesic relaxes the constant speed constraint of geodesics by allowing a diffeomorphism on the parameter $t$:
Thus if $\gamma(t)$ is a geodesic with constant speed, $\bar\gamma(u)=\gamma(u(t))$ is a pregeodesic for a diffeomorphism $u(t)$.
When considering singly parametric models ($m=1$), pregeodesics are simply intervals (i.e., $\bar\gamma_{p_0p_1}(u)=p_0+ u(p_1-p_0)$).
Some Riemannian manifolds have simple pregeodesics but geodesics mathematically complex to express.
For example, the pregeodesics of a Riemannian Klein hyperbolic manifold are Euclidean line segments~\cite{nielsen2012hyperbolic} making it handy for calculations.
See Algorithm~\ref{algo:multerrorpregeo} for the guaranteed Fisher-Rao approximation algorithm relying on pregeodesics instead of geodesics.


\begin{algorithm}
\KwIn{Probability distributions $p_{\theta_0}$ and $p_{\theta_1}$ on $\calM=\{p_\theta\}$}
\KwIn{Closed-form  pre-geodesic $\bar\gamma_{p_{\theta_0}p_{\theta_1}}(u)$}
\KwIn{Maximum multiplicative error $\eps>0$ }
\KwOut{A guaranteed approximation $\tilde\rho_\eps(\theta_0,\theta_1)\leq (1+\eps)\rho(\theta_0,\theta_1)$}
\tcc{Get lower bound $\rho_L$ and upper bound $\rho_U$ on the Fisher-Rao distance}
$\rho_L=\mathrm{LowerBound}(\theta_0,\theta_1)$\;
$\rho_U=\mathrm{UpperBound}(\theta_0,\theta_1)$\;
\If{$\frac{\rho_U}{\rho_L}>1+\eps$}{
\tcc{Cut the geodesic into two parts not necessarily balanced}
$m=\frac{\theta_0+\theta_1}{2}$\;
\Return $\mathrm{ApproxFisherRao}({\theta_0},m,\eps)+ \mathrm{ApproxFisherRao}(m,{\theta_1},\eps)$\; }\Else{\Return $\rho_U$}
\caption{Recursive algorithm $\mathrm{ApproxFisherRao}({\theta_0},\theta_1,\eps)$ for getting a $(1+\eps)$-approximation of the Fisher-Rao distance when only the pregeodesics $\bar\gamma(u)$ are known.}
\Mylabel{algo:multerrorpregeo}
\end{algorithm}


\subsection{Guaranteed approximations of Fisher-Rao distances  with arbitrary absolute errors}\Mylabel{sec:adderr}
Once we have a $(1+\eps)$-multiplicative factor approximation 
 algorithm $\mathrm{ApproxFisherRao}(\theta_0,\theta_1,\eps)$ of the Fisher-Rao distance $\rho(\theta_0,\theta_1)$, we can design a
 $(1+\delta)$ additive factor approximation 
 $\mathrm{ApproxAddErrFisher}(\theta_0,\theta_1,\eps)$
 for any $\delta>0$ (i.e., $\rho(\theta_0,\theta_1)\leq \tilde\rho_\delta(\theta_0,\theta_1)\leq \rho(\theta_0,\theta_1)+\delta$) as follows:
 Fix $\eps_0$ (say, $\eps_0=1$) and compute $\tilde\rho_{\eps_0}(\theta_0,\theta_1)$.
 We deduce from
 $$
 \rho(\theta_0,\theta_1)\leq  \tilde\rho_{\eps_0}(\theta_0,\theta_1)\leq (1+\eps_0)\,\rho(\theta_0,\theta_1)
 $$
 that $\rho(\theta_0,\theta_1)>\frac{1}{1+\eps_0}\tilde\rho_{\eps_0}(\theta_0,\theta_1)$.
 Now, choose $\eps>0$ so that $\epsilon\rho(\theta_0,\theta_1)\leq \delta$:
 $$
\eps\leq \frac{\delta}{\rho(\theta_0,\theta_1)}\leq \frac{(1+\eps_0)\delta}{\tilde\rho_{\eps_0}(\theta_0,\theta_1)}.
 $$
 Hence, we fix $\eps=\frac{(1+\eps_0)\delta}{\tilde\rho_{\eps_0}(\theta_0,\theta_1)}$ 
 and call the $(1+\eps)$-multiplicative error approximation algorithm $\mathrm{ApproxFisherRao}(\theta_0,\theta_1,\eps)$.
The additive error approximation algorithm is summarized in Algorithm~\ref{algo:adderror}.

\begin{algorithm}
\KwIn{Probability densities $p_{\theta_0}$ and $p_{\theta_1}$}
\KwIn{Maximum additive error $\delta>0$ }
\KwOut{A guaranteed approximation $\tilde\rho_\delta(\theta_0,\theta_1)\leq \rho(\theta_0,\theta_1)+\delta$}
\tcc{Fix initial multiplicative error $\eps_0$ to any arbitrary value}
$\eps_0=1$\;
$\tilde\rho_{\eps_0}=\mathrm{FisherRaoApprox}(\theta_0,\theta_1,\eps_0)$\;
\tcc{adjust multiplicative error}
$\eps=\frac{(1+\eps_0)\delta}{\tilde\rho_{\eps_0}}$\;
\Return $\mathrm{ApproxFisher}(\theta_0,\theta_1,\eps)$\;
\caption{Algorithm $\mathrm{ApproxAddErrFisherRao}(\theta_0,\theta_1,\delta)$: Converting an approximation scheme with multiplicative error to an approximation scheme with additive error.}
\Mylabel{algo:adderror}
\end{algorithm}

\underline{Caveat}: To implement these approximation algorithms, we need to have both lower and upper bounds on the Fisher-Rao distances.
In particular, to get arbitrarily fine approximations, we need to have  lower and upper bounds which are tight at infinitesimal scales. 
The following sections consider cases where we can obtain such bounds on the Fisher-Rao distances.

\section{Tight upper bounds on Fisher-Rao distances: The case of Hessian metrics}\Mylabel{sec:HessianMetric}

\begin{figure}%
\centering

\includegraphics[width=0.75\columnwidth]{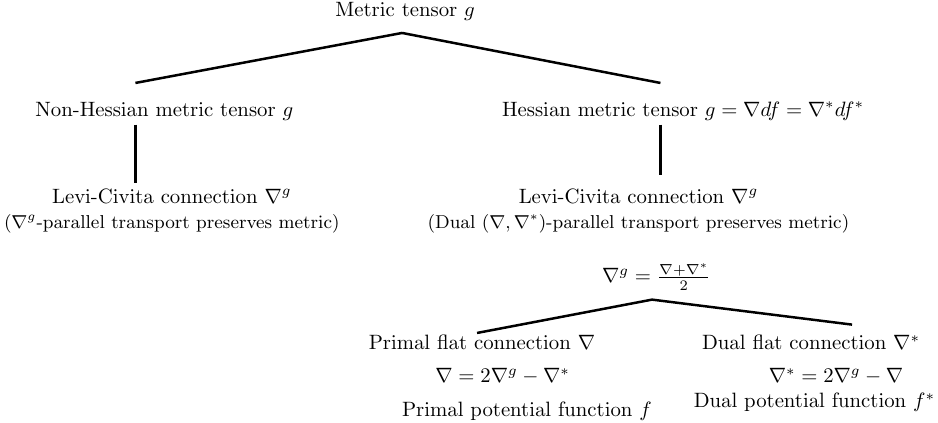}%

\caption{Diagram illustrating the canonical relationships between metric tensors $g$ and induced connections:
A metric-compatible Levi-Civita connection $\nabla^g$ is always induced by a metric $g$.
When the metric is Hessian, we can further associate to torsion-free flat connections $\nabla$ and $\nabla^*$ such that $\nabla^g=\frac{\nabla+\nabla^*}{2}$. }%
\Mylabel{fig:metricconnection}%
\end{figure}

A metric tensor $g$ induces a unique torsion-free metric connection called the Levi-Civita connection $\nabla^g$.
When the metric $g$ is Hessian, there exists two torsion-free flat connections $\nabla$ and $\nabla^*$ such that $g=\nabla d f=\nabla^* df^*$ for a pair of dual potential functions $f$ and $f^*$.
Furthermore, we have $\frac{\nabla+\nabla^*}{2}=\nabla^g$, i.e., the metric is preserved by dual parallel transport~\cite{IG-2016}.
Figure~\ref{fig:metricconnection} illustrates the canonical relationships between the metric tensor $g$ and induced connections.

Dually flat spaces in information geometry~\cite{ay2002dually,IG-2016} are special Hessian manifolds which admit global coordinate charts and dual potential functions which are Legendre-type convex functions when expressed in the dual $\nabla$- and $\nabla^*$-affine coordinate systems.
When the Fisher metric $g^{\Fisher}$ of a statistical model $\calM=\{p_\theta \st \theta\in\Theta\}$ is a Hessian metric, the Fisher-Rao manifold $(\calM,g^{\Fisher})$   is called a Hessian manifold~\cite{Shima-2007}.
When a Riemannian metric is Hessian,  
Recall that the Fisher metric is defined in the global chart by the Fisher information matrix $I(\theta)=-E_{p_\theta}\left[\nabla^2_\theta \log p_\theta(x)\right]$.
When the Fisher metric is Hessian, we have $G(\theta)=I(\theta)=\nabla^2 F(\theta)$ for a strictly convex and smooth potential function $F(\theta)$.
A dually flat space admits as a canonical divergence a Bregman divergence $B_F(\theta_1:\theta_2)=F(\theta_1)-F(\theta_2)-(\theta_1-\theta_2)^\top \nabla F(\theta_2)$ induced by a potential Legendre-type convex function $F(\theta)$ (defined up to a modulo affine terms), and two torsion-free flat connections $\nabla$ and $\nabla^*$ such that $\frac{\nabla+\nabla^*}{2}$ coincides with the Levi-Civita connection. 
These dual connections $\nabla$ and $\nabla^*$ coupled to the Hessian Fisher metric $g^\Fisher$ allows to connect any two points $p_{\theta_1}$ and $p_{\theta_2}$ on the manifold either by $\nabla$-geodesics $\gamma_{p_{\theta_1}p_{\theta_2}}$ or by $\nabla^*$-geodesics $\gamma_{p_{\theta_1}p_{\theta_2}}^*$ 
(Figure~\ref{fig:hessianfrmfd}).

\begin{figure}%
\centering

\includegraphics[width=0.75\columnwidth]{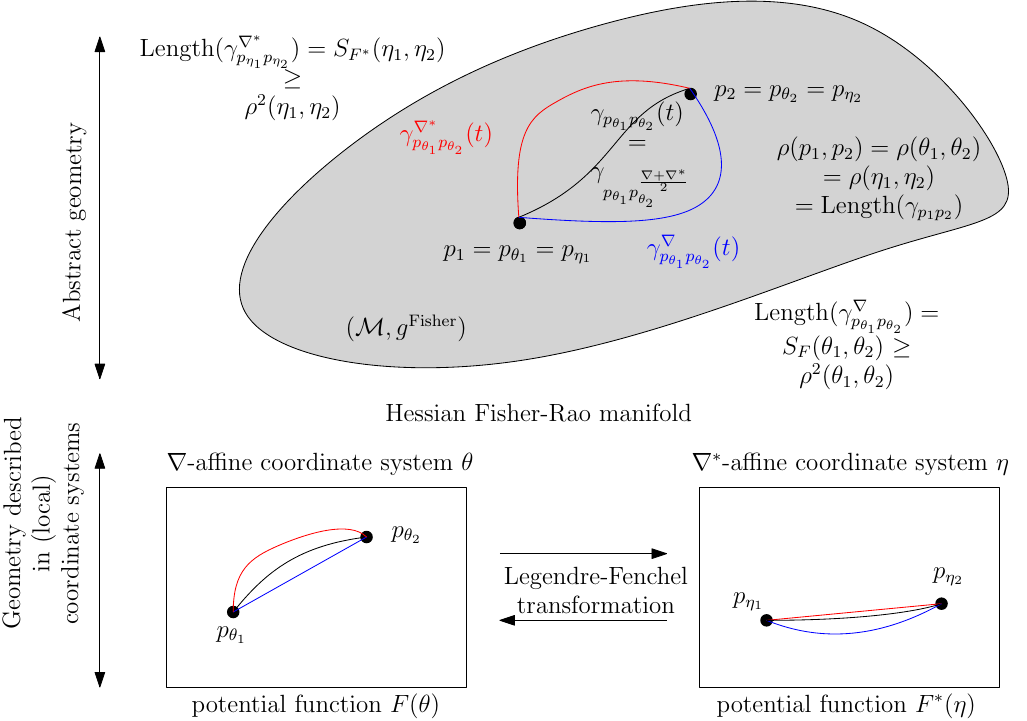}%

\caption{Hessian Fisher-Rao manifold: When the Fisher metric $g^\Fisher$ is Hessian, there exists two canonical torsion-free flat connections denoted by $\nabla$ and $\nabla^*$ which are conjugate to the Fisher metric  in a sense that their mid-connection $\frac{\nabla+\nabla^2}{2}$ coincides with the Levi-Civit\`a metric connection.
The Fisher-Rao length $\rho(p_1,p_2)$ is upper bounded by the square root of the Jeffreys-Bregman divergence which corresponds to the Fisher-Rao lengths of the $\nabla$-geodesic and the $\nabla^*$-geodesic. }%
\Mylabel{fig:hessianfrmfd}%
\end{figure}

The Jeffreys-Bregman divergence $S_F(\theta_1,\theta_2)$ is defined by
$$
S_F(\theta_1,\theta_2)=B_F(\theta_1:\theta_2)+B_F(\theta_2:\theta_1)=(\theta_2-\theta_1)^\top (\nabla F(\theta_2)-\nabla F(\theta_1)).
$$

We can interpret the Jeffreys Bregman divergence as the energy of dual geodesics~\cite{FisherRaoMVN-Nielsen-2023}:

\begin{proposition}[Theorem 3.2 of~\cite{IG-2016}]\Mylabel{prop:JBDRealization}
The Jeffreys-Bregman divergence $S_F(\theta_1,\theta_2)$ is interpreted as the energy induced by the Hessian metric $\nabla^2 F(\theta)$ on the dual geodesics:
$$
S_F(\theta_1,\theta_2)=\int_0^1 \ds^2(\gamma(t))\dt=\int_0^1 \ds^2(\gamma^*(t))\dt.
$$ 
\end{proposition}

\begin{proposition}[\cite{FisherRaoMVN-Nielsen-2023}]\Mylabel{prop:ubhessian}
The Fisher-Rao distance $\rho(\theta_1,\theta_2)$ of a statistical model $\{p_\theta\}$ with Hessian Fisher metric is upper bounded by the Jeffreys-Bregman divergence 
$S_F(\theta_1,\theta_2)=(\theta_2-\theta_1)\cdot(\nabla F(\theta_2)-\nabla F(\theta_1))$:
\begin{equation}\Mylabel{eq:frub}
\rho(\theta_1,\theta_2)\leq \sqrt{S_F(\theta_1,\theta_2)}.
\end{equation}
\end{proposition}

We have
$$
S_F(\theta_1,\theta_2)=S_{F^*}(\eta_1,\eta_2),
$$
where $F^*(\eta)$ is the dual potential function obtained from the Legendre transform. We have $\eta=\nabla F(\theta)$, $\theta=\nabla F^*(\eta)$, and ${F^*}^*=F$.

When $\theta_2\rightarrow \theta_1$, we have $\eta_2\rightarrow \eta_1$, and we get
\begin{equation}
S_F(\theta_1,\theta_2)=(\theta_2-\theta_1)\cdot(\eta_2-\eta_1) \rightarrow \ds^2(\theta_1,\theta_2-\theta_1).
\end{equation}
Thus the upper bound of Eq.~\ref{eq:frub} in tight at infinitesimal scale.

Many statistical models yield Fisher metrics:
For example, exponential families~\cite{barndorff2014information}, mixture families~\cite{IG-2016}, all statistical models of order $m=2$~\cite{UniElliptical-Mitchell-1988,burbea1988information,armstrong2015pontryagin}, and some elliptical distribution families~\cite{UniElliptical-Mitchell-1988,Elliptical-Mitchell-1989}, etc.
In general, there are obstructions for a 
  $m$-dimensional Riemannian metric $g$ to be Hessian~\cite{CurvatureHessian-2014}. 
Thus it is remarkable that exponential and mixture families of {\em  arbitrary order} yield Hessian Fisher metrics. 
	
One can test whether a  metric $g$ given in a coordinate system $\theta$ by $G(\theta)=[g_{ij}(\theta)]$ is Hessian or not as follows:	
	
\begin{property}[Hessian metric test~\cite{Shima-2007}]\Mylabel{prop:hessiantest}
A $m$-dimensional Riemannian metric $g$ expressed in the $\theta$-coordinate system is Hessian if and only if:
\begin{equation}\Mylabel{eq:HessianCheck}
\forall i,j,k\in\{1,\ldots,m\},\quad \frac{\partial g_{ij}(\theta)}{\partial\theta_k}=\frac{\partial g_{ik}(\theta)}{\partial\theta_j}.
\end{equation}
\end{property}

Beware that this property holds for a considered parameterization $\theta$ of the metric $g$. 

In general, when we change the coordinates of $\lambda$ to $\theta$, we consider the following covariant transformation rule of the metric tensor:
\begin{equation}\label{eq:HFIM}
g_\theta(\theta)=\Jac_\theta\lambda^\top\, g_\lambda(\lambda(\theta))\, \Jac_\theta\lambda,
\end{equation}
where $\Jac$ denote the Jacobian matrix: $\Jac_\theta\lambda=\left[\frac{\lambda_i}{\theta_j}\right]_{ij}$.
Thus to check that a metric $g(\lambda)$ given in some parameterization $\lambda$ is Hessian or not, we need to find a parameterization $\theta(\lambda)$ such that 
$g_\theta(\theta)=\Jac_\theta\lambda^\top\, g_\lambda(\lambda(\theta))\, \Jac_\theta\lambda=\nabla^2 F(\theta)$.

In 1D, $G(\theta)=[g_{11}(\theta)]>0$ and thus $g_{11}(\theta)=F_{g}''(\theta)$ for $F_g(\theta)=(\int_{y_0}^{\theta}\int_{x_0}^y g_{11}(x)\dx\dy)=a (\int\int g_{11}(\theta))+b$ for any real constants $x_0$, $y_0$, $a$ and $b$, where $\int\int g_{11}(\theta)$ denote the antiderivative of the antiderivative of $g_{11}$.

\begin{property}[1d Fisher metrics are Hessians]
All 1D Fisher metrics are Hessians.
\end{property}
Note that there are topological obstructions for the existence of dually flat structure $(g,\nabla,\nabla^*)$ on manifolds~\cite{ay2002dually}.
One way to obtain Hessian metrics is to choose a strictly convex and smooth function $F(\theta)$ and define $g$ in the $\theta$-coordinate system 
as $[g_{ij}(\theta)]=\nabla^2 F(\theta)$.

We may accelerate the Hessian metric test by finding a parameterization $\xi$ that yields the Fisher information matrix diagonal.
This is not always possible for find such orthogonal parameterizations $\xi$~\cite{huzurbazar1950probability}.

\begin{example}
 When the Bregman generator $F(\theta)$ is separable, i.e.,
$F(\theta)=\sum_{i=1}^m F_i(\theta_i)$
 for $m$ scalar Bregman generators $F_i$'s, we get $G(\theta)=\diag(F_1''(\theta_1),\ldots, F_m''(\theta_m))$ and the 
corresponding Riemannian metric is Hessian.
We get the following Riemannian distance~\cite{RiemannBregman-2019}:
\begin{equation}
\rho(\theta,\theta')=\sqrt{\sum_{i=1}^m \left(h_i(\theta_i)-h_i(\theta_i')\right)^2},
\end{equation}
where $h_i(x)=\int_c^x \sqrt{F_i''(u)}\, \du$.

Geometrically, $\rho$ is the Euclidean distance expressed in the $(h_1(\theta_1),\ldots, h_n(\theta_n))$ coordinates.
The Euclidean geodesics can be rewritten as $\gamma(t;\theta,\theta')=(\gamma_1(t;\theta_1,\theta'_1),\ldots, \gamma_n(t;\theta_1,\theta'_1))$ with
$$
\gamma_i(t;\theta_i,\theta_i')=h^{-1}\left((1-t)h_i(\theta_i)+t h_i(\theta_i')\right).
$$
That is, the Euclidean geodesics can be expressed using Kolmogorov-Nagumo quasi-arithmetic means~\cite{Kolmogorov-1930,Nagumo-1930}.
\end{example}

\begin{remark}
A 2D Hessian metric $g$ can always be expressed by a diagonal matrix in some coordinate system by a change of parameterization:
Indeed, let $\theta=(\theta_1,\theta_2)$ and $\eta=(\eta_1,\eta_2)$ be the dual affine $\nabla$- and $\nabla^*$-coordinate systems of the flat connections
$\nabla$ and $\nabla^*$. 
The Fisher metric is expressed in $\theta$ as $g(\theta)=\nabla^2 F(\theta)$ and in $\eta$ as $g(\eta)=\nabla^2 F^*(\eta)$.
Then we can consider the mixed parameterizations $\xi=(\theta_1,\eta_2)$ or $\xi'=(\eta_1,\theta_2)$ and obtain diagonal matrices for the Fisher information matrices with respect to those mixed parameterizations~\cite{IG-2016,miura2011introduction}.
\end{remark}

\begin{remark}
Consider the Fisher metric $g^\Fisher$ for the Cauchy family~\cite{nielsen2020voronoi} of distributions $\calC=\{p_{l,s}(x)=\frac{s}{\pi(s^2+(x-l)^2)} \st (l,s)\in\bbR\times\bbR_{>0}\}$: $g^\Fisher$ is expressed in the $(l,s)$-coordinates using  the Fisher information matrix is $G(l,s)=\frac{1}{2\,s^2}\, I$, where $I$ is the $2\times 2$ identity matrix.
The Amari-Chentsov $\alpha$-connections $\nabla^\alpha$ with
$\supleft{\alpha}\Gamma_{ijk}=\bar\Gamma_{ij}^k-\frac{\alpha}{2}T_{ijk}$ and $T_{ijk}=\partial_i\partial_j\partial_k \log p_\theta$ (cubic tensor)
all coincide with the Levi-Civita connection $\bar\nabla$ which is non-flat~\cite{UniElliptical-Mitchell-1988}.
Thus no pair of $\pm\alpha$-connections are dually flat.
However, the Fisher metric is Hessian since 2D~\cite{armstrong2015pontryagin}:
There exists a parameterization~\cite{nielsen2020voronoi} $\theta=(2\pi\frac{l}{s},-\frac{\pi}{s})$ such that the Fisher information matrix $I(\theta)=\nabla^2 F(\theta)$ for
 $F(\theta)=-\frac{\pi^2}{\theta_2}-\frac{\theta_1^2}{4\theta_2}-1$.
Thus we get dual flat torsion-free affine connections induced by $F(\theta)$ and its Legendre convex conjugate $F^*(\eta)$~\cite{IG-2016} for 
$F^*(\eta)=1-2\pi\sqrt{\eta_2-\eta_1^2}$ with $\eta=(l,l^2+s^2)$, see~\cite{nielsen2020voronoi}.
Notice that neither the Fisher information matrices $I(\theta)=\nabla^2 F(\theta)$ nor $I(\eta)=\nabla^2 F^*(\eta)$ are diagonal~\cite{nielsen2020voronoi}.
\end{remark}

\section{Tight lower bounds on Fisher-Rao distances from isometric embeddings}\Mylabel{sec:lowerboundiso}

Although there are many ways to design upper bounds on the Fisher-Rao distance, it seems more difficult to design lower bounds.
The classic technique for designing a lower bound on the Fisher-Rao distance of a statistical model $\calM$ is find an isometric embedding of $\calM$ as a submanifold of a higher-dimensional manifold for which the Riemannian distance is known in closed form.

We first illustrate this principle for the Fisher-Rao distance on the categorical distributions in~\S\ref{sec:isoSimplexHellinger} (multinomial distributions with a single trial also called multinoulli distributions by analogy to the link of Bernoulli distributions with Binomial distributions), and then discuss the technique for submodels of the family of multivariate normal distributions.

\subsection{Isometric embedding of the Fisher-Rao multinoulli manifold}\Mylabel{sec:isoSimplexHellinger}

Let $\Delta_{d-1}=\{(p^1,\ldots,p^d)\st \sum p^i=1, p^i>0\}$ be the family of categorical distributions 
 on a discrete sample space $\calX$ with $|\calX|=d$ (a discrete exponential family of order $m=d-1$ which can be parameterized by $\eta=(p^1,\ldots,p^{d-1})$.
The Fisher metric on $\Delta_{d-1}$ is
\begin{eqnarray}
g_{p}(v_1,v_2) &=& \sum_{i=1}^d \frac{v_1^iv_2^i}{p^i}, \forall v_1,v_2\in T_p\cong\bbR^d,\\
&=&  \left(\sum_{i=1}^{d-1} \frac{v_1^iv_2^i}{p^i}\right) + \frac{v_1^dv_2^d}{1-\sum_{j=1}^{d-1}p_j}.
\end{eqnarray}

Consider the embedding of the $(d-1)$-dimensional $\Delta_d$
  onto the Euclidean positive orthant sphere $\mathcal{S}_{d-1}^+\subset\bbR^{d}$ of radius $2$.
The ``square root mapping'' $f:\Delta_d\rightarrow \mathcal{S}_{d_1}^+$ with  
$$
\bar p=f(p)=2(\sqrt{p^1},\ldots,\sqrt{p^{d}})
$$
 yields an isometric embedding of the 
 Fisher-Rao manifold of categorical distributions into the Euclidean submanifold $\mathcal{S}_{d-1}^+\subset\bbR^d$.

The geodesics on $\mathcal{S}_{d-1}^+$ are not geodesics of $\bbR^d$, and thus the Euclidean submanifold $\mathcal{S}_{d-1}^+$ is not totally geodesic.
The inverse mapping $f^{-1}:\mathcal{S}_{d-1}^+\rightarrow \Delta_{d-1}$ with
 $p=f^{-1}(\bar p)=\frac{1}{4}((\bar p^1)^2,\ldots,(\bar p^{d})^2)$ yields the Fisher-Rao geodesics in $\Delta_d$ from the great arc of circle geodesics of $\mathcal{S}_{d-1}^+$. 
Thus the Fisher-Rao distance for $\Delta_{d-1}$ is
\begin{equation}
\rho_{\Delta_{d-1}}(p,q)=\rho_{\mathcal{S}_{d-1}^+}(\bar p,\bar q)=2\, \arccos \left(\sum_{i=1}^{d}\bar p^i\bar q^i\right)=2\arccos \left(\sum_{i=1}^{d} \sqrt{p^i q^i}\right).
\end{equation}
The Riemannian distance in Euclidean subspace $\bbR^{d}_{>0}$ is $\rho(p^+,q^+)=\frac{1}{\sqrt{2}}\|p^+-q^+\|_2$, and yields a lower bound on the Fisher-Rao distance $\rho(p,q)$ since the submanifold $\mathcal{S}_{d-1}^+$ is not totally geodesic.
This distance 
$$
\rho_{\bbR^{d}_{>0}}(\bar p,\bar q)=\frac{1}{\sqrt{2}} \sqrt{\sum_{i=1}^{d} (\sqrt{p^i}-\sqrt{q^i})^2}=\sqrt{1-\sum_{i=1}^{d} \sqrt{p^iq^i}}=\rho_{\Hellinger}(p,q)
$$
 is the Hellinger distance in $\Delta_{d-1}$.

Define the Bhattacharyya  coefficient $C(p,q)=\sum_{i=1}^{d} \sqrt{p^iq^i}$.
Then we have $\rho_\Fisher(p,q)=2\arccos(C(p,q))$ and $\rho_{\Hellinger}(p,q)=\sqrt{1-C(p,q)}$, and we check that the Hellinger distance is a lower bound on the Fisher-Rao distance between categorical distributions:
$$
\rho_{\Hellinger}(p,q)\leq \rho_\Fisher(p,q),
$$
since $\sqrt{1-u}\leq 2\arccos(u)$ for $u\in (0,1]$.
That is, the Hellinger distance which corresponds to the Fisher-Rao distance in the ambiant manifold is less or equal than the Fisher-Rao distance in $\Delta_{d-1}$:
Hellinger distance is a lower bound on $\rho(p,q)$ obtained by an isometric embedding.
Figure~\ref{fig:frembeddingsimplex} illustrates this Fisher-Rao isometric embedding of $\Delta_{d-1}$ into  $\mathcal{S}_{d-1}^+$.

\begin{figure}%
\center
\includegraphics[width=0.8\columnwidth]{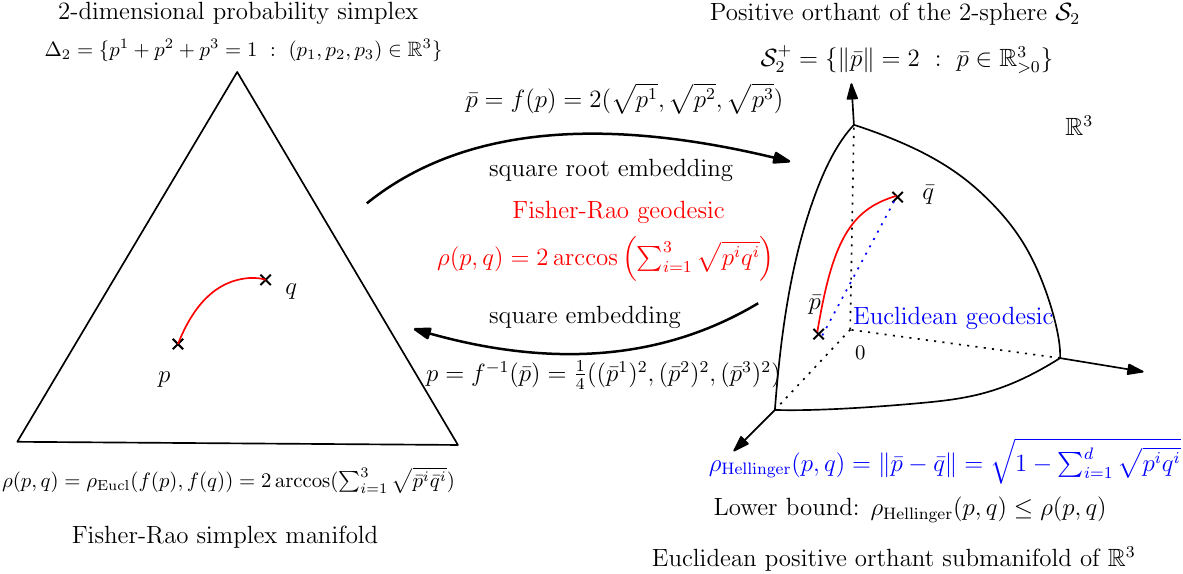}%
\caption{Fisher-Rao isometric embedding of the manifold of categorical distributions $\Delta_{d-1}$ into the Euclidean submanifold of the positive orthant sphere
 $\mathcal{S}_{d-1}^+\subset\bbR^{d}_{>0}$ of radius $2$. The submanifold $\mathcal{S}_{d-1}^+$ is not totally geodesic in $\bbR^{d}_{>0}$ and therefore the geodesic distance in  $\mathcal{S}_{d-1}^+$ is greater than the geodesic distance in $\bbR^{d}_{>0}$.}%
\Mylabel{fig:frembeddingsimplex}%
\end{figure}

Using Nash embedding theorem~\cite{nash1954c1,han2006isometric,gromov2017geometric}, a $m$-dimensional Riemannian manifold $(M,g)$ can always be embedded as a submanifold of the Euclidean manifold $(E,g_E)$ of dimension
$n \leq 2m$. 
Let $f:M\rightarrow E$ be the embedding function.
Then we have
$$
\rho_g(p_1,p_2) \geq \rho_E(f(p_1),f(p_2))=\|f(p_1)-f(p_2)\|_2.
$$ 
Let us notice that computing explicitly the embedding function $f$ is a difficult problem in general although some algorithmic works tackle this problem~\cite{zhong2018computing}. 
Hilbert~\cite{hilbert1901ueber} proved that the hyperbolic upper plane manifold (Poincar\'e hyperbolic metric for $m=2$) cannot be fully embedded in the $n=3$-dimensional Euclidean space. Only a partial band of the upper plane can be isometrically embedded onto the pseudo-sphere~\cite{pressley2010hyperbolic} of $\bbR^3$.

\subsection{Case of totally geodesic submanifolds}\Mylabel{sec:totallygeo}
We may embed a Fisher-Rao manifold $(\calS,g_\calS^\Fisher)$ into a higher-dimensional Fisher-Rao manifold $(\calM,g_\calM^\Fisher)$ by a mapping $f:\calS\rightarrow\calM$.
The mapping $f$ is isometric if the metric of the embedded manifold $\bar\calS=\{f(p) \st p\in \calS\}$ viewed as a submanifold of $\calM$  restricted to $\bar\calS$ coincide with the Fisher-Rao metric of $\calS$.
For a totally geodesic submanifold $\bar\calS\subset\calM$, the geodesics passing through two points $\barp_0$ and $\barp_1$ of $\bar\calS$ fully stay in $\bar\calS\in\calM$  (Figure~\ref{fig:totallygeodesic}).

\begin{property}[Comparing Fisher-Rao distances under isometric embededings]
When  an isometric embedding $f:\calS\rightarrow\calM$ is totally geodesic, the Fisher-Rao distances coincide: 
$\rho_{\calS}(\theta_0,\theta_1)=\rho_{\calM}(\bar\theta_0,\bar\theta_0)$ where $\bar\theta=f(\theta)$.
When the isometric embedding is not totally geodesic, then we get the Fisher-Rao distance in $\calM$ provides a lower bound on the Fisher-Rao distance in $\calS$:
$\rho_{\calM}(\bar\theta_0,\bar\theta_0)\leq \rho_{\calS}(\theta_0,\theta_1)$.
\end{property}

For example, consider the $m=\frac{d(d+3)}{2}$-dimensional Fisher-Rao manifold of $d$-variate normal (MVN) distributions  $\calM=\{p_{\mu,\Sigma}(x) \st (\mu,\Sigma)\in \bbR^d\times\bbP(d)\}$.
We have the Fisher-Rao length element given by
$$
(\ds^\Fisher_\calM)^2 = \dmu^\top \Sigma^{-1} \dmu + \frac{1}{2}\tr\left(\left(\Sigma^{-1}\dSigma\right)^2\right).
$$

Let $\calS_\mu=\{q_\Sigma(x)=p_{\mu,\Sigma}(x) \st \Sigma\in \bbP(d)\}$ be the Fisher-Rao manifold of MVN distributions centered at position $\mu$ with Fisher-Rao distance denoted by 
$\rho_\mu$.
We have 
$$
{\ds^\Fisher_\calS}^2 = \frac{1}{2}\tr\left(\left(\Sigma^{-1}\dSigma\right)^2\right).
$$

Consider the isometric embedding $f_\mu:\calS_\mu\rightarrow\calM$ with $f_\mu(\Sigma)=(\mu,\Sigma)$.
Then $\bar\calS_\mu=\{f_\mu(\Sigma)\st \Sigma\in\bbP(d)\}$ is a totally geodesic submanifold of $\calM$, and we have
$$
\rho_{\calM}((\mu,\Sigma_0),(\mu,\Sigma_1))=\rho_\mu(\Sigma_0,\Sigma_1).
$$

\subsection{Case of non-totally geodesic submanifolds}\Mylabel{sec:nontotallygeo}

Let $\calS_\Sigma=\{q_\mu(x)=p_{\mu,\Sigma}(x)\st \mu\in\bbR^d\}$ be the Fisher-Rao manifold of MVN distributions with fixed covariance matrix $\Sigma$ and Fisher-Rao distance denoted by $\rho_\Sigma$.
Then $f_\Sigma:  \calS_\Sigma\rightarrow\calM$ with $f_\Sigma(\mu)=(\mu,\Sigma)$ yields an embedded submanifold 
$\bar\calS_\Sigma=\{f_\Sigma(\mu)=(\mu,\Sigma)\st \mu\in\bbR^d\}$ which is not totally geodesic.
Hence, we have the following lower bound on $\rho_{\calM}$:
$$
\rho_{\calM}((\mu_0,\Sigma),(\mu_1,\Sigma))\geq \rho_\Sigma(\mu_0,\mu_1).
$$

Many submanifolds of the MVN manifold have been considered in the literature and their totally geodesic properties or not have been investigated~\cite{pinele2020fisher}.

\begin{figure}%
\centering
\includegraphics[width=\columnwidth]{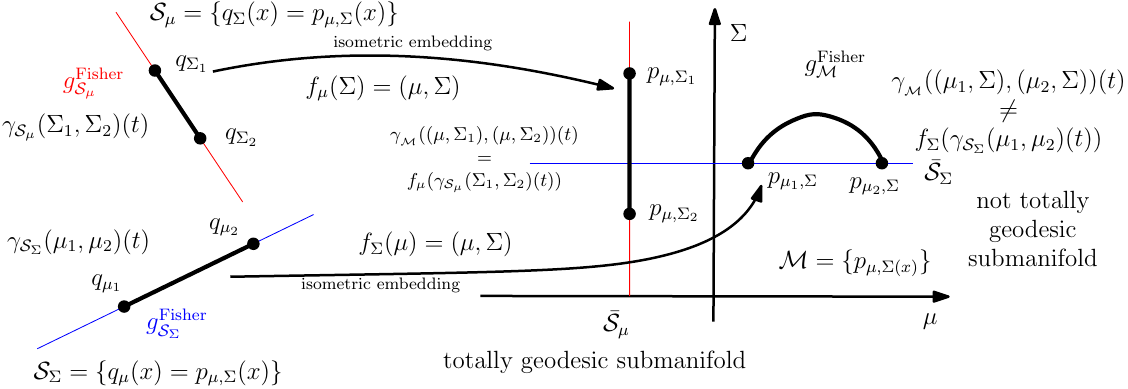}%
\caption{Totally versus non-totally geodesic submanifolds: The geodesics of a totally geodesic submanifold with respect to the ambiant metric fully stays on the embedded submanifold. }%
\Mylabel{fig:totallygeodesic}%
\end{figure}

\section{Approximating Fisher-Rao distances on elliptical distribution manifolds}\Mylabel{sec:EllDist}

\subsection{Fisher-Rao distances for univariate elliptical distribution families}\Mylabel{sec:frls}

A univariate elliptical family~\cite{UniElliptical-Mitchell-1988} $\El_h=\{p_{m,v}(x)=\frac{1}{\sqrt{v}} h(\Delta_v^2(x,m)) \st (m,v)\in\bbR\times\bbR_{>0}\}$ is defined according to a function $h$ where $\Delta_v^2(x,m)=
\frac{(x-m)^2}{v}$ is the 1d squared Mahalanobis distance. 
A location-scale family is defined by $\calM=\{p_{l,s}(x)=\frac{1}{s}\, p_\std\left(\frac{x-l}{s}\right) \st (l,s)\in \bbR\times\bbR_{>0}\}$ where $p_\std$ is the standard density corresponding to the parameter $(l=0,s=1)$.
Thus location-scale families can be interpreted as elliptical families in 1d with $(l,s)=(m,\sqrt{v})$ and $p_\std(u)=h(\sqrt{u})$.

The Fisher information matrix~\cite{UniElliptical-Mitchell-1988,burbea1988information} of a location-scale family  $\calM$
with continuously differentiable standard density $p_\std(x)$ defined on the full support $\calX=\bbR$ is
 $$
I(\lambda)=\frac{1}{s^2}\mattwotwo{a^2}{c}{c}{b^2},$$
 where 
\begin{eqnarray*}
a^2 &=& E_{p_\std}\left[\left(\frac{p'_\std(x)}{p_\std(x)}\right)^2\right],\\
b^2 &=& E_{p_\std}\left[\left(1+x\frac{p'_\std(x)}{p_\std(x)}\right)^2\right],\\
c &=& E_{p_\std}\left[\frac{p'_\std(x)}{p_\std(x)} \left(1+ x \frac{p'_\std(x)}{p_\std(x)} \right) \right].
\end{eqnarray*}

\begin{table}
\centering
\begin{tabular}{l|ll}
Family name & $a$ & $b$\\ \hline
$k$-Student & $\frac{k+1}{4(k+3)}$ & $\frac{3(k+1)}{4(k+3)}$\\
Normal $(k\rightarrow\infty)$& $\frac{1}{4}$ & $\frac{3}{4}$\\
Cauchy $(k=1)$ & $\frac{1}{8}$ & $\frac{3}{8}$
\end{tabular}
\caption{Some parameters characterizing families of univariate elliptical distributions.}
\Mylabel{tab:ell1d}
\end{table}

When the standard density function is even (i.e., $p(x)=p(-x)$), we get a diagonal Fisher information matrix ($c=0$) 
$$
I_{a,b}(l,s)=\frac{1}{s^2}\, \mattwotwo{4a}{0}{0}{4b-1}.
$$
Table~\ref{tab:ell1d} displays the values of $a$ and $b$ for the univariate normal distributions, Cauchy distributions, and Student distributions with $k$ degrees of freedom.

The Fisher information matrix can be reparameterized with 
$$
\theta(l,s)=\left(\frac{a}{b}l,s\right),
$$ 
so that the Fisher information matrix with respect to $\theta$
becomes 
$$
I_\theta(\theta)= {b^2} \, \mattwotwo{\frac{1}{\theta_2^2}}{0}{0}{\frac{1}{\theta_2^2}}.
$$ 
That is, the Fisher information matrix with respect to parameter $\theta$ is a conformal metric of the Poincar\'e metric of the  hyperbolic upper plane.
It follows that the Fisher-Rao geometry is hyperbolic with curvature
$\kappa=-\frac{1}{b^2}<0$, and that the Fisher-Rao distance between two densities $p_{l_1,s_1}$ and $p_{l_2,s_2}$ of a location-scale family is 
$$
\rho((l_1,s_1),(l_2,s_2))= b\, \rho_\Poincare\left(\left(\frac{a}{b}l_1,s_1\right),\left(\frac{a}{b}l_2,s_2\right)\right)
$$
 where 
$$
\rho_\Poincare(\theta_1,\theta_2)=\arccosh\left(1+\chi(\theta_1,\theta_2)\right),$$ 
where $\arccosh(u)=\log(u+\sqrt{u^2-1})$ for $u>1$ and 
$$
\chi((l_1,s_1),(l_2,s_2)):=\frac{(l_1-l_2)^2+(s_1-s_2)^2}{2s_1s_2}.
$$  

\begin{remark}
All Fisher metrics of 1d elliptical distributions are Hessian metrics:
Let $\xi(m,v)=(\frac{m}{v},-\frac{1}{2v})$ be a reparameterization of $(m,v)$.
Then the Fisher information matrix $I(\xi)$ can be expressed as a Hessian of a potential function (\cite{UniElliptical-Mitchell-1988}, p 9).
\end{remark}

\subsection{Multivariate elliptical distribution families}\Mylabel{subsec:elldist}
Let $(\bbR^d,\calB(\bbR^d),\mu)$ be the measure space with the $\sigma$-algebra Borel sets $\calB(\bbR^d)$ equipped with the  Lebesgue measure  $\mu$.
Let $X$ be a $d$-dimensional random variable following a continuous elliptical distribution~\cite{EllipticalDistribution-Kelker-1970,ED-Chmielewski-1981}.
A $d$-dimensional random vector  $X\sim\El_\Psi(m,V)$ follows an elliptical distribution~\cite{EllipticalDistribution-Kelker-1970,ED-Chmielewski-1981,ED-Portfolio-2016} (ED) if and only if its characteristic function (CF) $\varphi_X(t)$ 
(characterizing uniquely  the probability distribution) can be written as
$$
\varphi_X(t)=E[\exp(it^\top X)]=\exp(it^\top m)\, \Psi(t^\top V t),
$$
for some scalar function $\Psi(u)$ called the characteristic generator.
Parameter $m\in\bbR^d$ denotes the location parameter (or median) and parameter $V\in\bbP(\bbR,d)$ is the scale parameter where $\bbP(\bbR,d)$ indicates
the cone of symmetric positive-definite $d\times d$ matrices (with $\bbP(\bbR,d)\subset\Sym(\bbR,d)$ where $\Sym(\bbR,d)$ denotes the set of real symmetric matrices).
EDs yield a rich class of probability distributions including also heavy-tailed distributions.

In general, probability density functions (PDFs) of EDs may not exist (e.g., EDs with positive semi-definite $V$ also called equivalently non-negative definite) or may not be in closed form but characteristic functions always exist.

When the probability density function (PDF) exists, we denote $X\sim\El_{d,h}(m,V)$, and write the PDF in the following canonical form:
$$
p_{m,V}(x)=\frac{c_d}{\sqrt{\det(V)}}   \, h\left(\Delta_V^2(x,m)\right)
$$
where 
$$
\Delta_V^2(x,m)=(x-m)^\top V^{-1} (x-m),
$$ 
is the quadratic distance called the squared Mahalanobis distance~\cite{Mahalanobis-1936}.
 Let 
$$
\calE=\{p_{m,V}(x) \st (m,V)\in \Lambda=\bbR^d\times\bbP(\bbR,d)\}
$$
Let $\lambda=(m,V)$ denote the $m=d+\frac{d(d+1)}{2}=\frac{d(d+3)}{2}$ parameters of an elliptical distribution.

Elliptical distributions are symmetric distributions~\cite{MVSymmetric-2018}, radially symmetric around the location parameter $m$, with elliptically-contoured densities (i.e., densities are constant on ellipsoids). They are popular in applications since they include heavy-tailed distributions.

In general, the mean $E[X]$ and covariance matrix $\Cov[X]$ may not exist (eg., Cauchy or L\'evy distributions). 
However, when they exist, the covariance matrix is $cV$ for a constant   $c=-2\, \left.\frac{d}{\du}\Psi(u)\right|_{u=0}$. 

The two most popular EDs met in the literature are (1) the normal distributions and the generalized Gaussian distributions~\cite{MVExpPowerDistrib-2008,andai2009geometry,FisherKLDGenGaussian-2011,FisherRaoGenGaussian-2012,dytso2018analytical,KLD-MGGD-2019}, 
and (2) the Student $t$-distributions~\cite{MVT-2004} including the Cauchy distributions~\cite{fdivCauchy-2023} and the Laplace distributions~\cite{MVLaplace-2001}:
EDs~\cite{Muirhead-2009} also include Pearson type II and type VII distributions (including $t$-distributions), contaminated normal distributions, logistic distributions~\cite{yin2022new},  hyperbolic distributions~\cite{HyperbolicDistribution-1995,Schmidt-2003}, stable distributions, just to name a few.

For example, the Multivariate Generalized Gaussian Distributions(MGGDs) and Multivariate $t$-distributions (MTDs) are defined as follows:
\begin{itemize}
\item  
Multivariate Generalized Gaussian Distributions (MGGDs)~\cite{KLD-MGGD-2019} with shape parameter $k\in\bbN$ have PDFs:
\begin{eqnarray*}
p_{k,m,V}^\MGGD(x) &=& \frac{k \Gamma\left(\frac{d}{2}\right)}{\pi^{\frac{d}{2}} \Gamma\left(\frac{d}{2k}\right) 2^{\frac{d}{2k}}}
\, \frac{\exp\left(-\frac{1}{2} \left((x-m)^\top V^{-1} (x-m)\right)^k\right)}{\sqrt{\det(V)}},\\
&=& \frac{c_d}{\sqrt{\det(V)}} \,  \exp\left( -\frac{1}{2} \Delta^2(x,m)^{k} \right).
\end{eqnarray*}

MGGDs include the multivariate Gaussian distributions (or normal distributions)  when $k=1$.

\item 
Multivariate $t$-distributions~\cite{KLD-MTD-2023} (MTDs) with $k\in\bbN$ degrees of freedom have PDFs:
\begin{eqnarray*}
p_{k,m,V}^\MTD(x) &=& 
\frac{\Gamma\left(\frac{k+d}{2}\right)}{\Gamma\left(\frac{k}{2}\right) \, k^{\frac{d}{2}}\, \pi^{\frac{d}{2}}} 
\frac{\left(1+\frac{1}{d} (x-m)^\top V^{-1} (x-m)\right)^{-\frac{k+d}{2}}}{\sqrt{\det(V)}},\\
&=& \frac{c_d}{\sqrt{\det(V)}} \,   \left(1+\frac{1}{d}\Delta^2(m,x)\right)^{-\frac{k+d}{2}}.
\end{eqnarray*}

MTDs include the Cauchy distributions ($k=1$) and in the limit case of $k\rightarrow\infty$ the normal distributions.

\end{itemize}

\subsection{Fisher-Rao metric and Fisher-Rao geodesics}

The Fisher information matrix of EDs was first studied in~\cite{UniElliptical-Mitchell-1988} for univariate EDs and in~\cite{Elliptical-Mitchell-1989} for multivariate EDs.
Let $V=L^\top L$ be the Cholesky decomposition of the scale matrix $V$, and define the random variable 
$$Z=L^{-1}(X-m)\sim \El_{d,c_d,h}(0,I),
$$ 
following the standard elliptical distribution,
and define 
$$
W=\frac{d\log h(\|Z\|^2)}{d(\|Z\|^2)}.
$$

Then the length element $\ds$ of the Fisher metric $g^\Fisher_{a,b}$ can be expressed as
\begin{equation}\Mylabel{eq:metric}
\ds^2 = 4a\, \dm^\top V^{-1} \dm + 2b\, \tr((V^{^1}\dV)^2) + \frac{4b-1}{4}\, \tr^2(V^{-1}\dV),
\end{equation}
where
$a=\frac{1}{d} E[\|Z\|^2 W^2]$, 
and  $b=\frac{1}{d(d+2)} E[\|Z\|^4 W^2]$.

In particular, we have
$
a_\MGGD=
k^2\, \frac{ \Gamma\left(2+ \frac{d}{2k}-\frac{1}{k}\right)}{ 2^{\frac{1}{k}}\, d\, \Gamma\left(\frac{d}{2k}\right)}
 , \quad  b_\MGGD=\frac{d+2k}{4d+8}$
and
$a_\MTD=b_\MTD=\frac{k+d}{4(k+d+2)}$.

The geodesic equation~\cite{FisherRao-Elliptical-1997,EllipticalDistance-CalvoOller-2002,EllipticalGeodesic-Inoue-2015} for elliptical distributions is given by the following second-order ODE:

$$
\left\{
\begin{array}{lll}
\ddotm-\dV V^{-1}\dotm = 0\\
\ddotV+\alpha\, \dotm\dotm^\top - \beta\, \dotm^\top V^{-1} \dotm V - \dotV V^{-1} \dotV =0
\end{array}
\right.,
$$
with $\alpha=\frac{a}{b}$ and $\beta=\frac{a(4b-1)}{(8+4d)b^2-db}$.

In general, the Fisher geodesics of EDs are not known in closed-form~\cite{MVRao-2005} except when the location parameter is fixed~\cite{FisherRao-Elliptical-1997}, and  for the multivariate normal case with initial conditions~\cite{Eriksen-1987} and bounding conditions~\cite{Kobayashi-2023}. 

When the location parameter $m$ is prescribed, the ODE becomes 
$$
\ddotV - \dotV V^{-1} \dotV =0,
$$ 
which solves as:

\begin{itemize}
\item with initial conditions $(V_0,S_0\in\Sym(\bbR,d))$:

\begin{equation}
V_t= V_0^{\frac{1}{2}}\, \exp\left(t\, V_0^{-\frac{1}{2}}\, S_0\, V_0^{-\frac{1}{2}} \right)\, V_0^{\frac{1}{2}}, \quad t\in\bbR.
\end{equation}

More generally, we have~\cite{EllipticalDistance-CalvoOller-2002}
$\dotV(t)=V(t)H$ and $V(t)=V_0\exp(Ht)$, where $H$ is a constant matrix satisfying $VH=H^\top V$ (e.g., the identity matrix).

\item with boundary conditions $V_0$ and $V_1$:

$$
V_t=V_1^{\frac{1}{2}}\, \exp\left(t\, \log(V_0^{-\frac{1}{2}}\, V_1\, V_0^{-\frac{1}{2}}) \right)\, V_1^{\frac{1}{2}}, \quad t\in[0,1].
$$
\end{itemize}

It is remarkable that the geodesics does not depend on the elliptical generator $h$ of the elliptical family.

Since $\calE'_m$ is a totally geodesic submanifold of $\calE$, we get in closed-form the Fisher-Rao geodesics for scale ED subfamilies, and the Fisher-Rao distance can be calculated in closed form~\cite{FisherRao-Elliptical-1997}.

\subsection{Calvo and Oller's Fisher-Rao isometric embeddings onto the SPD matrix cone}\Mylabel{sec:CO}

We report the isometric embeddings introduced by Calvo and Oller~\cite{EllipticalDistance-CalvoOller-2002} of the Fisher-Rao elliptical manifolds $\calS$ into the Fisher-Rao manifold of centered normal distributions 
$\calM=\{p_{0,\Sigma}(x)\st \Sigma\in\bbP(d+1)\}$.

We consider the following Fisher-Rao isometric diffeomorphisms $f_{\alpha,\beta}: \Lambda \rightarrow \bbP(d)$ defined by:
$$
\barE=f_{\alpha,\beta}(E)=(\det(V))^\alpha\, \mattwotwo{V+\beta mm^\top}{\beta m}{\beta m^\top}{1},
$$
where $\alpha=\alpha(a,b)$ and $\beta=\beta(a,b)$ are scalar parameters depending on $a$ and $b$, see~\cite{EllipticalDistance-CalvoOller-2002} .
For $\theta=(m,V)$, let $\bar\theta=f_{\alpha,\beta}(m,V)$ be the $(d+1)\times (d+1)$ positive-definite matrix parameter.
The embedded submanifold $\bar\calS=\{f_{\alpha,\beta}(m,V)\st (m,V)\in\bbR^d\times\bbP(d)\}$ is not totally geodesic in $\calM$, so we get a lower bound on the Fisher-Rao distance:
$$
\rho((m_0,V_0),(m_1,V_1))\geq \rho_{\calM}(\bar\theta_0,\bar\theta_1).
$$

Thus we get from this isometric embedding a lower bound on the Fisher-Rao distance between elliptical distributions which is tight at infinitesimal scale:

\begin{property}[Calvo-Oller lower bound]\Mylabel{prop:FRBLB}
The Fisher-Rao distance $\rho((m_0,V_0),(m_1,V_1))$ between two elliptical distributions $\El_{d,h}(m_0,V_0)$ and $\El_{d,h}(m_1,V_1)$  is lower bounded by the SPD cone distance on the $\bar P=f_{\alpha,\beta}(m,V)$ embedding:
$$
\rho((m_0,V_0),(m_1,V_1)) \geq  \rho_{\SPD}(\barP_0,\barP_1). 
$$
\end{property}

We may project the SPD geodesic in $\bbP(d+1)$ onto the submanifold $\bar\calS$ and pullback that projected curve onto the elliptical manifold in a manner similar to~ what has been done for MVNs in~\cite{FisherRaoMVN-Nielsen-2023}.
The Fisher orthogonal projection of $P\in\bbP(d+1)$ onto $\bar\calS$   can be implemented using results described in the appendix of~\cite{EllipticalDistance-CalvoOller-2002}). Then we may approximate the Fisher-Rao length of that pullback curve using discretization described in~\S\ref{sec:approxFRlength}.

Notice that the Fisher-Rao geodesics with boundary conditions in $\calS$ are not known in closed-form~\cite{MVRao-2005} in general, except for the special case of multivariate normal distributions~\cite{Kobayashi-2023}.

Next, we propose a new metric distance for elliptical distribution families which is fast to compute and an approximation of the Fisher-Rao distance based on pulling back geodesics of that new distance.

\subsection{The Birkhoff-Calvo-Oller distance for elliptical distribution families}\Mylabel{sec:elliptical}

\subsubsection{Birkhoff projective distance on the SPD matrix cone}

The {\em Birkhoff projective distance}~\cite{Birkhoff-1957,CTP-2021} on the symmetric-positive definite (SPD) matrix cone $\bbP(d)$ is defined by
\begin{eqnarray*}
\rho_\Birkhoff(P_0,P_1)
&=& \log\left(\frac{\lambda_{\max}(P_0^{-\frac{1}{2}}P_1 P_0^{-\frac{1}{2}})}{\lambda_{\mmin}(P_0^{-\frac{1}{2}}P_1 P_0^{-\frac{1}{2}})}\right),\\
 &=& \log\left(\frac{\lambda_{\mmax}(P_0^{-1}P_1)}{\lambda_{\mmin}(P_0^{-1}P_1)}\right),
\end{eqnarray*}
where $\lambda_{\mmin}(M)$ and $\lambda_{\mmax}(M)$ denote the smallest and largest eigenvalue of matrix $M$, respectively.

The Birkhoff  distance is symmetric and satisfies the triangular inequality.
However, we have $\rho_\Birkhoff(P_0,P_1)=0$ if and only if $P_0=\lambda P_1$ for some $\lambda>0$.
Thus the Birkhoff  distance  is called a {\em projective distance} as it measures distances between rays of the SPD cone.
Since its coincides with the Hilbert log-cross ratio distance on any section of the SPD cone, the Birkhoff projective distance has also been called the Hilbert projective distance~\cite{lemmens2014birkhoff}.

Let us define our new distance based on the Calvo-Oller isometric embedding:

\begin{definition}[Birkhoff-Calvo-Oller distance]\Mylabel{def:FBdist}
The Birkhoff-Calvo-Oller distance between two elliptical distributions $\El_{d,h}(m_0,V_0)$ and $\El_{d,h}(m_1,V_1)$ with Fisher metric $g_{a,b}^\Fisher$ 
(Eq.~\ref{eq:metric}) is
\begin{equation}
\rho_\Birkhoff((m_0,V_1),(m_1,V_1)) = \log \frac{\lambda_{\mmax}(M_{01})}{\lambda_{\mmin}(M_{01})},
\end{equation}
where $\barP_i=f_{a,b}(m_i,V_i)$ is the Calvo-Oller embedding of parameters into the higher-dimensional SPD cone,
and $\lambda_\mmin$ and $\lambda_\mmax$ the smallest and largest eigenvalues of matrix $M_{01}= \barP_0^{-\frac{1}{2}}\barP_1 \barP_0^{-\frac{1}{2}})$.
\end{definition}

\begin{property}
The Birkhoff-Calvo-Oller distance is a metric distance.
\end{property}
Birkhoff distance is projective on the SPD cone. But we have $\bar P_0=\lambda\bar P_1$ only for $\lambda=1$, hence the Birkhoff-Calvo-Oller distance is a metric distance.
In practice, we compute approximately these extreme smallest and largest eigenvalues using the power method iterations.
See also~\cite{mostajeran2023differential} for distances with underlying Finsler geometry based on extreme eigenvalues.
 
\begin{figure}%
\centering  
\begin{tabular}{cc}
\includegraphics[width=0.30\textwidth]{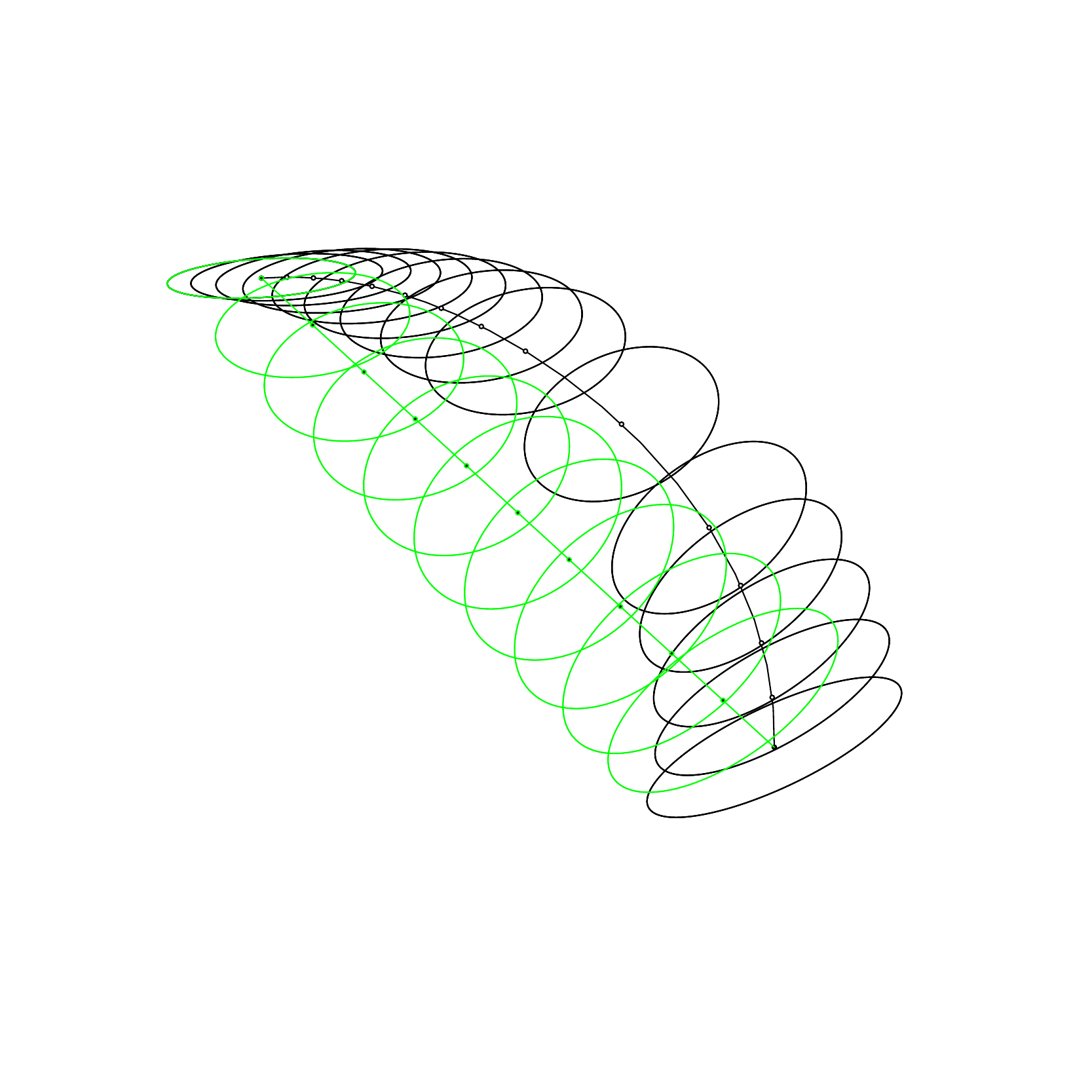}%
&
\includegraphics[width=0.30\textwidth]{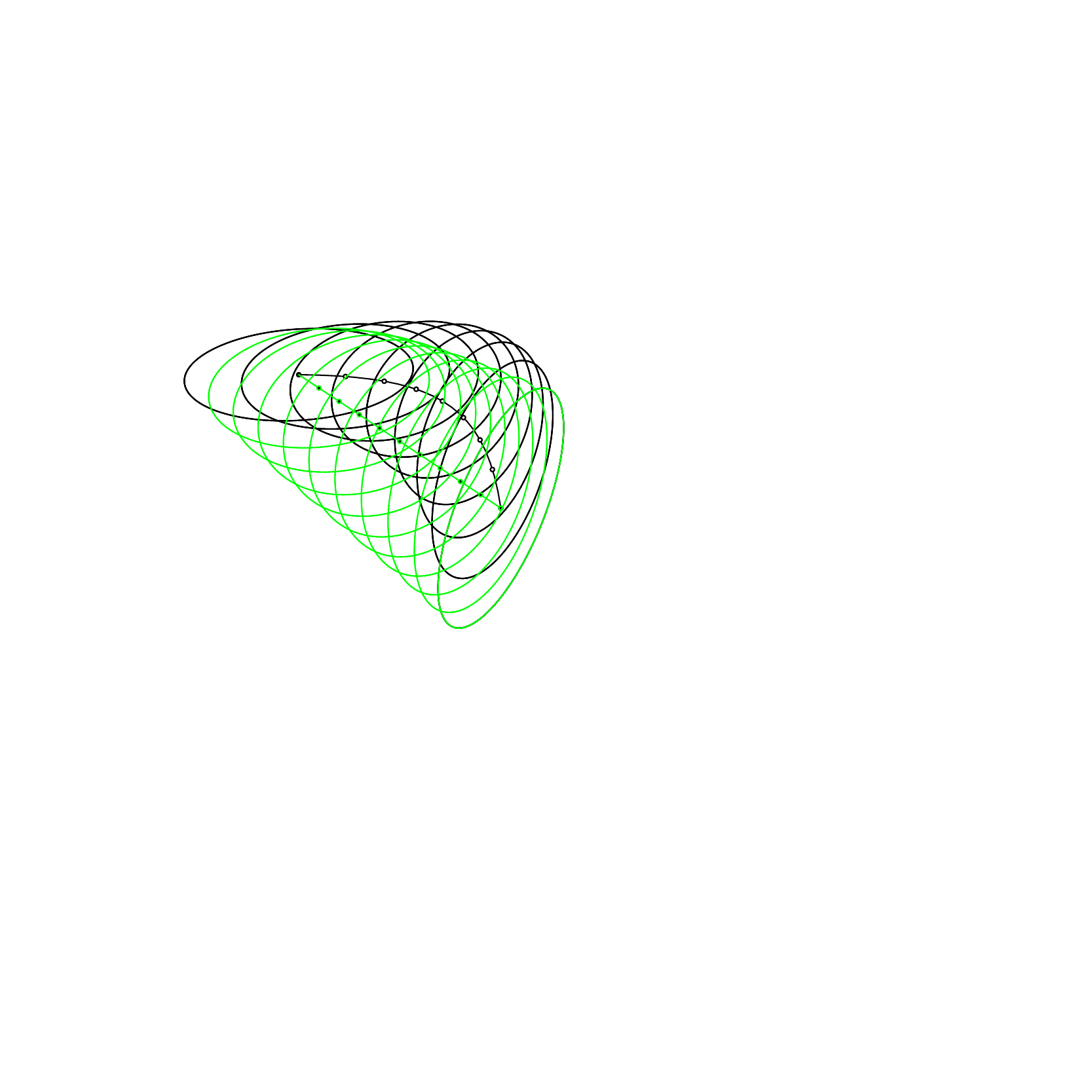}\\
\includegraphics[width=0.30\textwidth]{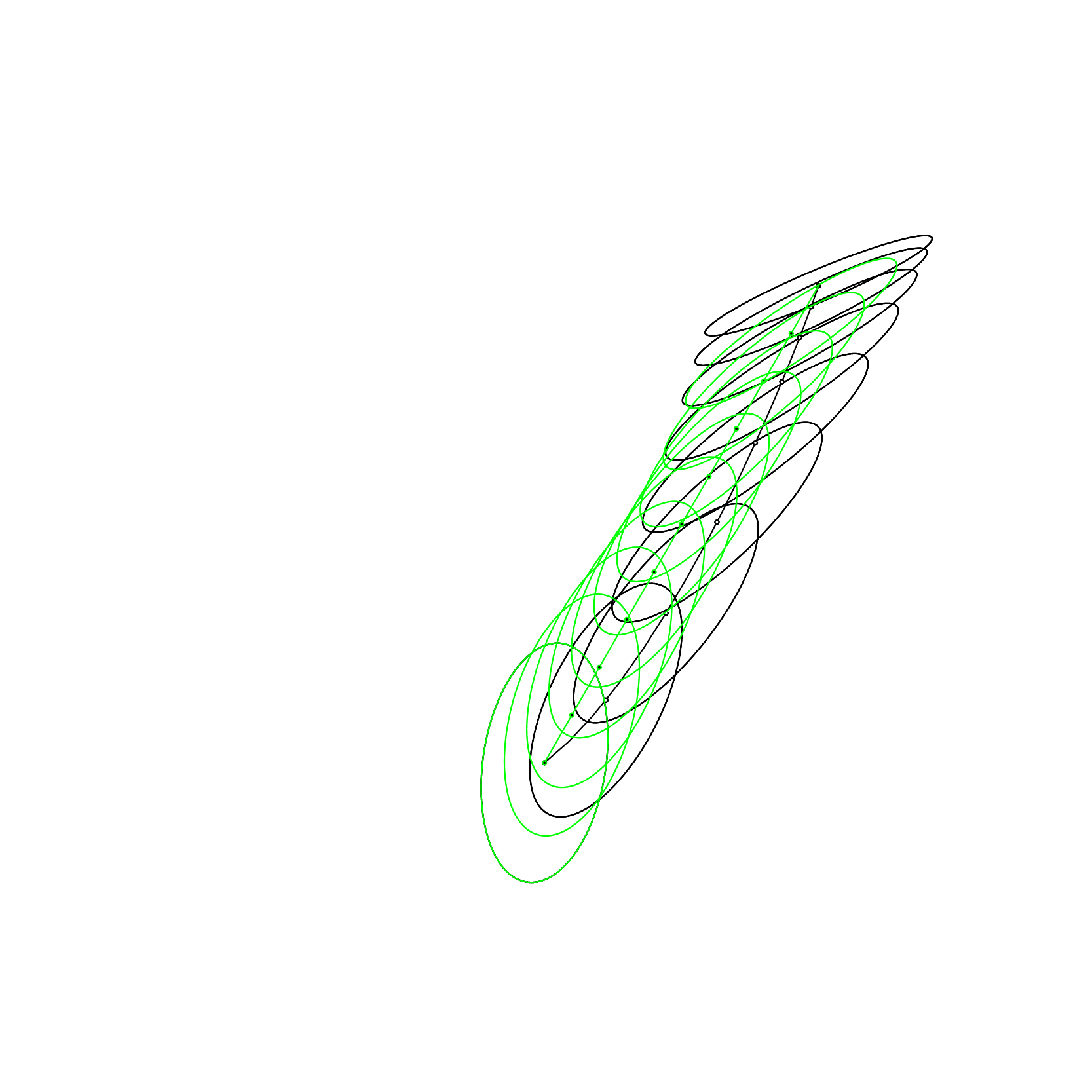}%
&
\includegraphics[width=0.30\textwidth]{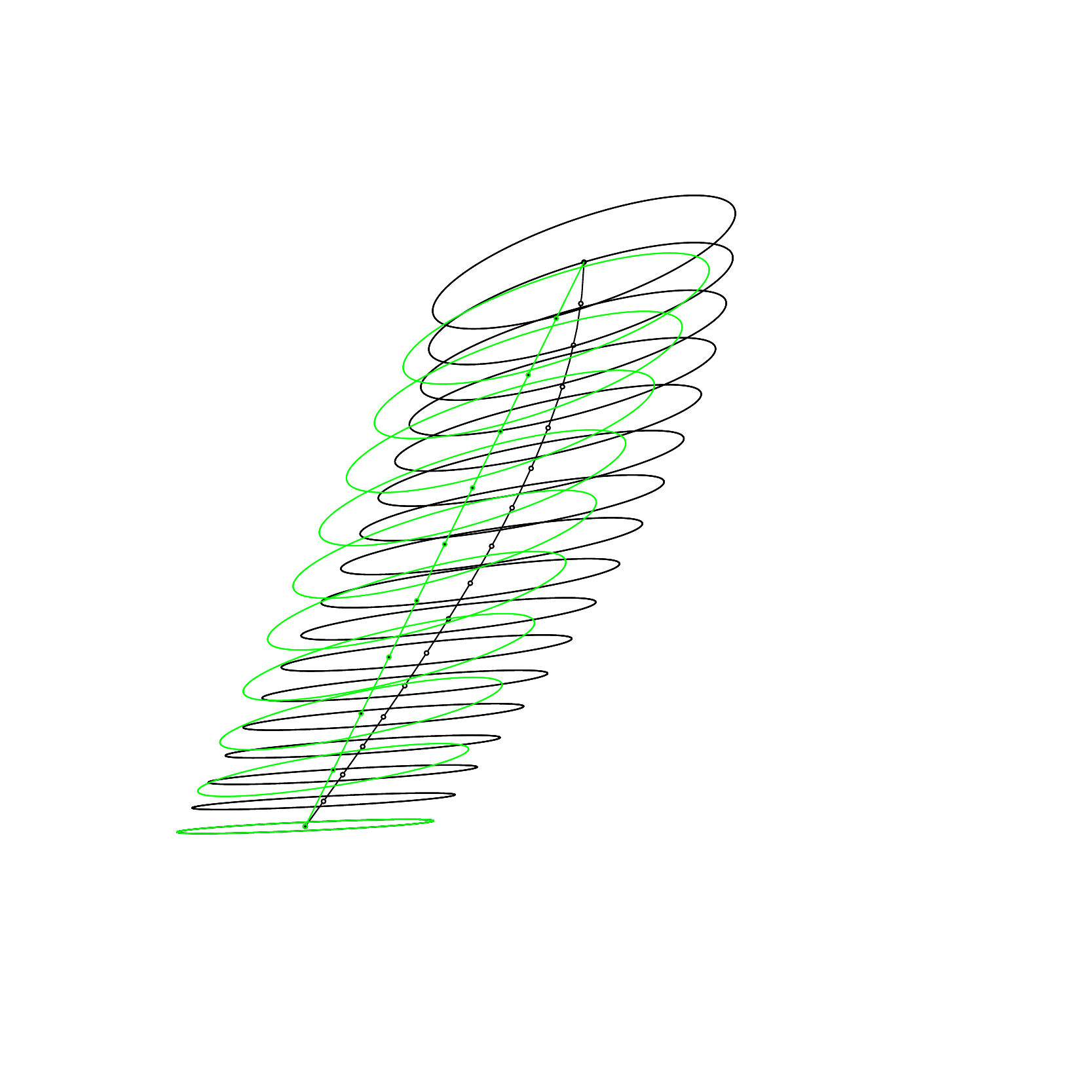}
\end{tabular}
\caption{Some examples of Fisher-Rao geodesics (black) and Birkhoff pulled back curves (green) between pairs of bivariate normal distributions.}%
\Mylabel{fig:Hilbertgeodesic}%
\end{figure}

\subsubsection{Approximations of the Fisher-Rao distances by pulling back Birkhoff geodesics}\Mylabel{sec:frhilbertgeo}

The Birkhoff geometry on the SPD cone allows us to define Birkhoff geodesics which can be projected on $\bar\calS$ and then pulled back on the Fisher-Rao elliptical manifolds using the inverse mapping  $f_{\alpha,\beta}^{-1}$ of the embedding diffeomorphisms $f_{\alpha,\beta}$ of Calvo and Oller~\cite{CalvoOller-1991}.
Geodesics $\gamma_\Birkhoff(P_0,P_1)$ in the Birkhoff SPD cone are straight lines~\cite{nussbaum1994finsler} parameterized as follows:
$$
\gamma_\Birkhoff(P_0,P_1;t):=\left(\frac{\beta\alpha^t-\alpha\beta^t}{\beta-\alpha}\right)P_0+
\left(\frac{\beta^t-\alpha^t}{\beta-\alpha}\right)P_1,
$$
where $\alpha=\lambda_\mmin(P_1^{-1}P_0)$ and $\beta=\lambda_\mmax(P_1^{-1}P_0)$.

Figure~\ref{fig:Hilbertgeodesic} displays some Fisher-Rao geodesics between pairs of multivariate normal distributions obtained from~\cite{Kobayashi-2023}, and their corresponding pulled back curves.

\section{Maximal invariant and Fisher-Rao distances}\Mylabel{sec:maxinvardist}

In this section, we take an algebraic point of view to find closed-form expressions of Fisher-Rao distances using the framework of group actions and maximal invariants.

\subsection{Group action and maximal invariant}

A statistical distance $D(p_{\theta_1},p_{\theta_2})$ between two statistical distributions $p_{\theta_1}$ 
and $p_{\theta_2}$ of a statistical model $\calM=\{p_\theta\st\theta\in\Theta\}$  (not necessarily a metric distance) is equivalent to a distance between their parameters:
$D_\calM(\theta_1,\theta_2)=D(p_{\theta_1},p_{\theta_2})$.
Thus a distance can be interpreted as a function $D_\calM(\cdot,\cdot)$ on the product manifold $\calM\times\calM$.
In information geometry, those distances are called divergences, contrast functions, or yokes.
Let us write for short $D(\theta_1,\theta_2)=D_\calM(\theta_1,\theta_2)$.

A group $\calG=(\calX,\mathrm{op},e)$ consists of a set $\calX$ equipped with a binary operation $\mathrm{op}$ and a neutral element $e\in\calX$ with respect to the group operation.
For example, $(\bbR,+,0)$ or $(\bbR_{\not=0},\times,1)$ are the additive group of real numbers  
and the multiplicative group of  real numbers, respectively.

Consider a group $\calG$ acting on the pair of parameters $(\theta,\theta')\in\Theta\times\Theta$ while preserving the distance function:
$$
\forall g\in \calG,\quad D_\calM(g.(\theta_1,\theta_2))=D_\calM(\theta_1,\theta_2),
$$
where $g.(\theta,\theta')$ denotes the group action of the element $g\in\calG$ on $(\theta,\theta')\in\Theta\times\Theta$.
We can visualize the group action of $g$ on an element $(\theta_1,\theta_2)$ by its orbit:
$$
O_{\theta_1,\theta_2}=\{g.(\theta_1,\theta_2)\st g\in \calG\},
$$
which has constant distance value.

Whenever two distinct orbits $O_{\theta_1,\theta_2}$ and $O_{\theta_1',\theta_2'}$ yield different function (distance) values 
$D({\theta_1,\theta_2})\not = D({\theta_1',\theta_2'})$, we say that the distance function is a {\em maximal invariant distance} (Figure~\ref{fig:maxinvar}).

\subsubsection{Maximal invariant under translation and 1D Euclidean distance}

\begin{example}
For example, consider the 1D Euclidean distance: $D_E(\mu_1,\mu_2)=|\mu_1-\mu_2|$ (or $D_E(\mu_1,\mu_2)=(\mu_1-\mu_2)$ assuming $\mu_1\geq \mu_2$)
 and the group $\calG=\{g\in\bbR\}$ with neutral element $0$ and group action the addition.
Consider the translation operation as the group action:
$g.(\mu_1,\mu_2)=(\mu_1+g,\mu_2+g)$.
We have $D_E(g.(\mu_1,\mu_2))=D_E(\mu_1+g,\mu_2+g)=\mu_1+g-(\mu_2-g)=\mu_1-\mu_2=D_E(\mu_1,\mu_2)$.
So the 1D Euclidean distance is invariant under the translation action of the group $\calG=(\bbR,+,0)$.
Moreover, $D_E(\mu_1,\mu_2)\not=D_E(\mu_1',\mu_2')$ when $\mu_1-\mu_2\not=\mu_1'-\mu_2'$.
Thus the translation is a maximal invariant for the 1D Euclidean distance.
\end{example}

Consider the family of 1D squared Mahalanobis distances: 
$$
\Delta_{\sigma^2}^2(\mu_1,\mu_2)= \frac{(\mu_1-\mu_2)^2}{\sigma^2},
$$
  defined for any $\sigma>0$.
The  distances $\Delta_{\sigma^2}$ for $\sigma>0$ are invariant under translation (group action): $\Delta_{\sigma^2}^2(\mu_1+l,\mu_2+l)=\Delta_{\sigma^2}^2(\mu_1,\mu_2)$.

Eaton's theorem~\cite{eaton1989group} states that any invariant function is a function of a maximal invariant.
Thus we have:
$$
\Delta_{\sigma^2}(\mu_1,\mu_2)=f_\sigma(D_E(\mu_1,\mu_2)).
$$
In our case, we get $f_\sigma(u)=\frac{u}{\sigma}$.

\begin{figure}%
\centering
\includegraphics[width=0.55\columnwidth]{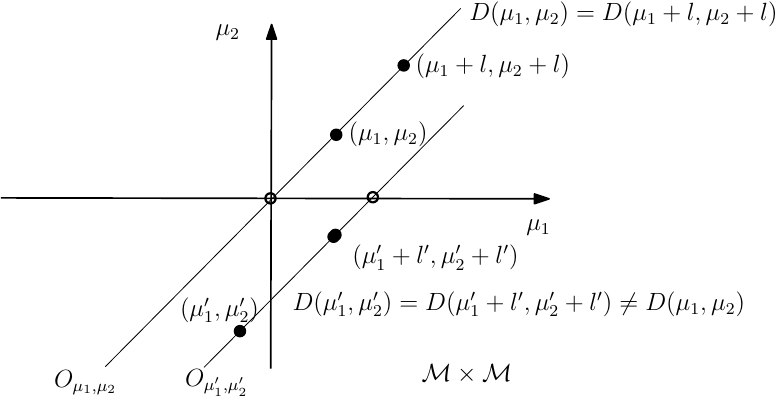}%
\caption{A maximal invariant for the distance function on the product manifold under a group action (here, translation) is such that different group orbits have distinct distance values.}%
\Mylabel{fig:maxinvar}%
\end{figure}

Now, consider the $f$-divergences between two densities of a location family.
We have $D_f(\mu_1,\mu_2)=I_f(p_{\mu_1}:p_{\mu_2})$ which is invariant under translation when the standard density is an even function: $p(-x)=p(x)$:
$$
D_f(g.\mu_1,g.\mu_2)=D_f(\mu_1,\mu_2).
$$

Thus using Eaton's theorem, we necessarily have the existence of functions $h_f$ such that
$I_f(p_{\mu_1}:p_{\mu_2})=h_f(D_E(\mu_1,\mu_2))$.
The advantage of this algebraic method is that although $I_f(p_{\mu_1}:p_{\mu_2})$ may not be in closed form (i.e., the Jensen-Shannon divergence between two isotropic Gaussian distributions), we can nevertheless use the {\em structural result} that $I_f(p_{\mu_1}:p_{\mu_2})=h_f(D_E(\mu_1,\mu_2))$ to deduce that
$$
I_f(p_{\mu_1}:p_{\mu_2})>I_f(p_{\mu_3}:p_{\mu_4}) \Leftrightarrow D_E(\mu_1,\mu_2)>D_E(\mu_3,\mu_4).
$$
For example, we can compare exactly the Jensen-Shannon divergence between isotropic Gaussians although no closed-form formula is known.
See~\cite{fdivCauchy-2023,nielsen2023f} for more advanced examples of use of maximal invariant in statistical dissimilarities. 

In general, when the maximal invariant is $t$-dimensional~\cite{nielsen2023f}, we can express the invariant function using $t$-terms.
This structural knowledge helps use to define syntax formula trees which may be searched using symbolic regression~\cite{makke2024interpretable}.

\subsubsection{Maximal invariant under rescaling and 1D hyperbolic distance}

Now, consider the rescaling operation as a group action of the group $\calG=(\bbR,\times,1)$.
Then the 1D hyperbolic distance $D_H(s_1,s_2)=|\log\frac{s_1}{s_2}|$ is a maximal invariant for the group action $g.s=g\lambda s$ (Figure~\ref{fig:maxinvarscale}).
Consider the scale family $\calM=\{p_s(x)=\frac{1}{s} f(\frac{x}{s})\}$ for a standard density $f_\std(x)$.
The $f$-divergences are invariant under rescaling: $I_f(p_{s_1}:p_{s_2})=I_f(s p_{s_1}:s p_{s_2})$.
Thus using Eaton's theorem, we have $I_f(p_{s_1}:p_{s_2})=h_f(D_H(s_1,s_2))$ for some function $h_f$.
Therefore the Fisher-Rao length element is invariant under rescaling,
 and thus the Fisher-Rao distance is also invariant under rescaling.
Hence the Fisher-Rao distance between two scale distributions can be expressed as a function of the 1D hyperbolic distance.

For example, 
\begin{itemize}
\item The Fisher-Rao distance between two distributions of the exponential distribution family $\calM=\{p_\lambda(x)=\lambda \exp(-\lambda x)\}$ (a scale family with $f_\std(x)=\exp(-x)$ and $s=\frac{1}{\lambda}$ on the support $\bbR_{>0}$)
is expressed as
$$
\rho_{\mathrm{Exponential}}(p_{\lambda_1},p_{\lambda_2})=D_H(\lambda_1,\lambda_2).
$$
That is $h_\FR(u)=u$.

\item The Fisher-Rao distance between two distributions of the Rayleigh distribution family $\calM=\{p_{\sigma^2}(x)=\frac{x}{\sigma^2} \exp(-x^2/(2\sigma^2))\}$
(a scale family with $f_\std(x)=x\exp(-x^2)$ and $s=\sigma^2$ on the support $\bbR_{>0}$)
is expressed as
$$
\rho_{\mathrm{Rayleigh}}(p_{\sigma_0^2},p_{\sigma_1^2})= D_H(\sigma_0^2,\sigma_1^2).
$$
That is $h_\FR(u)=u$.
\end{itemize}

\begin{remark}
Notice that the Mahalaobis distance $\Delta_\Sigma(\mu_1-\mu_2)=\sqrt{(\mu_1-\mu_2)^\top\, \Sigma^{-1}\ (\mu_1-\mu_2)}$ can be interpreted as a 1D
 Mahalanobis distance $\Delta_\Sigma(\mu_1-\mu_2)=\Delta_1(0,\Delta_\Sigma(\mu_1-\mu_2))$.
Using this observation and by a change of variable, under conditions on the standard density of a $d$-variate location family, 
we can show that $f$-divergences between two $d$-variate distributions of an elliptical family amounts to a scalar function of their Mahalanobis distance~\cite{nielsen2024f}.
\end{remark}

\begin{figure}%
\centering
\includegraphics[width=0.4\columnwidth]{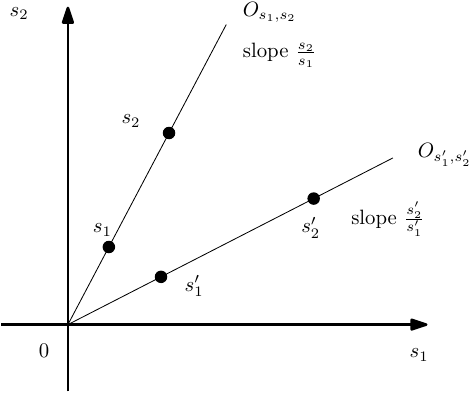}%
\caption{A maximal invariant distance under a scaling group action is such that different group orbits have distinct distance values.}%
\Mylabel{fig:maxinvarscale}%
\end{figure}

Many exponential families admit group actions~\cite{barndorff1982exponential}.
Spherical distributions are orthogonally invariant~\cite{eaton1986characterization}.	
We refer the reader to~\cite{tojo2021harmonic} for a generic constructing of exponential families invariant under transformation groups using  representation theory.

\subsection{Positive affine group action on families of elliptical distributions}

We check that the length element $\ds^\Fisher$ of Eq.~\ref{eq:metric} is invariant under affine transformations:

$$
\dm'=A\dm,\quad \dV'=A^{-\top}\dV A^{-1}.
$$

$$
(A^{-1})^\top=(A^\top)^{-1}
$$
We use the cyclic property of the trace: $\tr(ABC)=\tr(BCA)=\tr(CAB)$.

$$
\ds^2_{m,V} = 4a\, \dm^\top V^{-1} \dm + 2b\, \tr((V^{^1}\dV)^2) + \frac{4b-1}{4}\, \tr^2(V^{-1}\dV)
$$

Notice that we also have invariance of the Mahalanobis distance under affine transformations~\cite{Elliptical-Mitchell-1985}:
$$
\Delta_{AVA^\top}(Am_0+a,Am_1+a)=\Delta_{V}(m_0,m_1).
$$

More generally, the invariance under affine transformations hold for generalized skew-elliptical distributions~\cite{SkewEDAffine-2018}.
It is known that all the affine-invariant metrics on the SPD cone can be written canonically as~\cite{AffineInvariantMetricSPD-Thanwerdas-2019}:
$$
g_V(S_1,S_2)=\alpha\tr\left(V^{-1}S_1\, V^{-1}S_2\right) + \beta \tr\left(V^{-1}S_1\right)\,\tr\left(V^{-1}S_2\right),
$$
with $\alpha,\beta>0$ and $\frac{\beta}{\alpha}>-\frac{1}{d}$.
The corresponding infinitesimal length element is
$$
\ds_V(\dS)=\alpha\tr\left((V^{-1}\dS)^2\right) + \beta \tr^2\left(V^{-1}\dS\right).
$$
	
Calvo and Oller mentioned the	maximal invariant for multivariate normal distributions in~\cite{CalvoOller-1991} which allows to reduce the number of arguments in the  distance from $\frac{d(d+3)}{2}$ to $1+\frac{d(d+1)}{2}$, see~\cite{FisherRaoMVN-Nielsen-2023}.

\section{Conclusion: Summary and some open problems}\Mylabel{sec:concl}

The Fisher-Rao distance is a natural metric distance between probability distributions of a parametric statistical model~\cite{dowty2018chentsov}. 
However, its computational tractability has limited its use in practice (e.g., lack of closed-form formula for multivariate normal distributions~\cite{Kobayashi-2023}).
In this work, we have considered several approximation and bounding techniques for the Fisher-Rao distance.

First, we reported (i) a structured formula for the Fisher-Rao distance of uniparametric model (Proposition~\ref{prop:fr1d}), (ii) extended this formula for stochastically independent product of statistical models (Proposition~\ref{prop:FRindepsto}), and (iii) designed a canonical Fisher-Manhattan upper bound on Fisher-Rao distances
 (Proposition~\ref{prop:FisherManhattan}).

Second, we considered generic schemes to approximate Fisher-Rao distances either by approximating the Fisher-Rao lengths of curves with explicit parameterization 
(Proposition~\ref{prop:approxFRlength}). When Fisher-Rao geodesics are available in closed-form, we designed a simple technique to approximate the Fisher-Rao distance based on the metric property of Riemannian geodesics  (Proposition~\ref{prop:frepsgeo}). 
Furthermore, when Fisher-Rao geodesics or pregeodesics are available in closed forms, we can guarantee arbitrarily finely the approximation provided we have both tight lower and upper bounds on the Fisher-Rao distance (Algorithm~\ref{algo:multerrorgeo}, Algorithm~\ref{algo:multerrorpregeo}, and Algorithm~\ref{algo:adderror}).

When the Fisher-Rao metric $g$ is Hessian  we get a canonical pair of dual torsion-free flat affine connections $\nabla$ and $\nabla^*$ which mid-connection $\frac{\nabla+\nabla^*}{2}$ coincides with the Levi-Civita metric connection induced by the Fisher metric.
In  dimension $1$ and $2$, all analytic metrics are Hessian metrics~\cite{armstrong2015pontryagin}.
We noticed that the dually flat connections $\nabla$ and $\nabla^*$ associated to a Hessian metric may not be necessarily  Amari's expected $\alpha$-connections as this can be attested for some univariate elliptical models~\cite{UniElliptical-Mitchell-1988} like the family of Cauchy distributions for which all Amari's expected $\alpha$-connections coincide with the non-flat Fisher-Rao connection.
 Yet the Fisher Cauchy metric is Hessian and admits two canonical flat connections~\cite{nielsen2020voronoi}. 
To implement Algorithm~\ref{algo:adderror} for Fisher Hessian metrics, we reported a canonical upper bound on  their Fisher-Rao distances (Proposition~\ref{prop:ubhessian}), and show how to design lower bounds from isometric embeddings in \S\ref{sec:lowerboundiso}.
In particular, we considered Calvo and Oller isometric embeddings~\cite{EllipticalDistance-CalvoOller-2002} for multivariate elliptical distribution families and discussed how to implement the various approximation and upper bounds for those families in \S\ref{sec:elliptical}.
We also proposed a fast alternative metric distance called the Birkhoff-Calvo-Oller distance for multivariate elliptical distribution families
(Definition~\ref{def:FBdist}). 

Finally, we took an algebraic approach to study Fisher-Rao distances from the angle of maximal invariants of distance functions defined on product manifolds in \S\ref{sec:maxinvardist}.
Provided that dissimilarity measures are invariant under the action of a group, we seek to express those distances in terms of maximal invariants.
For example,  many distances between elliptical densities like the Fisher-Rao distance, $f$-divergences, and Wasserstein distances~\cite{WassersteinED-Muzellec-2018}  are invariant under the group action of the positive affine group.
When the scale/covariance matrices are fixed, the Fisher-Rao distance and $f$-divergences can be expressed as a function of their Mahalanobis distances~\cite{nielsen2024f}.
This results is useful in practice because it allows one to compare exactly those distances even when not available in closed-form by doing an equivalent comparison on their closed-form Mahalanobis distances.

For future work, we further expect fast and numerically robust  algorithms for approximating the Fisher-Rao distances based on differential-geometric ingredients (e.g., Riemannian exponential/logarithmic maps, geodesics with initial conditions in closed-form, etc) 
and efficient implementations in software packages~\cite{le2023parametric}.  

Finally, let us list some open problems:
\begin{description}

\item[Problem 1.] Can we obtain a {\em closed-form formula} for the Fisher-Rao distance between multivariate normal distributions?
Geodesics with boundary value conditions have recently been elicited~\cite{Kobayashi-2023}.

	\item[Problem 2.] Characterize the classes of statistical models which are guaranteed to have {\em unique} Fisher-Rao geodesics.

		\item[Problem 3.] Given a Riemannian metric $g$ expressed using parameterization $\lambda$ (i.e., given $g(\lambda)$ on a single global chart), how to find, if possible, a parameterization $\theta$ and a potential convex function $F(\theta)$ such that $g(\theta)=\nabla^2 F(\theta)$ (i.e., check whether the geometric metric $g$ is Hessian or not). See~\cite{ciaglia2017hamilton,ciaglia2019generalized}.
	
	\item[Problem 4.] Study $\calM=\{p_\theta:\theta\in\Theta\}$ as a {\em complex manifold} instead of a real manifold when the statistical model $m$ is of even order. 
	See preliminary work in~\cite{burbea1982entropy} and K\"ahler geometry in information geometry~\cite{ciaglia2017hamilton}.
	
		\item[Problem 5.] Study the geometry of {\em irregular statistical models} for which the Fisher information matrix  is not well-defined (i.e., either undefined or 
		not positive-definite).
		See preliminary work using Finsler geometry and the Hellinger divergence in~\cite{AmariFinsler-1984}.

\item[Problem 6.] When and how can we embed exponential families or statistical models of order $m$ into high-dimensional symmetric positive-definite matrix cones  
$\bbP(m')$ (with $m'>m$) equipped with a scaled trace metric? See potential related paper~\cite{chua2003relating}.
\end{description}

\vskip 0.5cm
\noindent Acknowledgments: We thank Kyle Cranmer for pointing out a ``typographic error'' (in Eq.~\ref{eq:FRcat}) in former versions of this manuscript.

\bibliographystyle{plain}
\bibliography{FisherRaoApproximationSchemeBIB}

\appendix
\section{Symbolic computing of Fisher information}\label{app:FIM}

We use the computer algebra software {\tt Maxima} (\url{https://maxima.sourceforge.io/}) for computing symbolically the Fisher information matrix.
For example, the code snippet below calculates the Fisher information of bivariate normal distributions~\cite{sato1979geometrical}:

\begin{verbatim}
kill(all)$
v:matrix([x],[y]);
mu:matrix([a],[b]);
Sigma: matrix([c,d],[d,e]);

assume(c>0);assume(e>0);
assume(determinant(Sigma)>0);

/* pdf of a bivariate normal distribution */
pdf:1/(2*%pi*sqrt(determinant(Sigma)))*exp(-(1/2)*transpose(v-mu).invert(Sigma).(v-mu));

/* Fisher information matrix */
hessian(-log(pdf),[a,b,c,d,e]);
integrate(integrate(pdf*%,x,minf,inf),y,minf,inf);
\end{verbatim} 

We can check that the expectation of the score vanishes:
\begin{verbatim}
/* check that the expectation of the score is zero */
derivative(log(pdf),a,1);
integrate(integrate(pdf*%,x,minf,inf),y,minf,inf);
\end{verbatim} 

The Fisher information matrix for some cases of trivariate normal distributions was reported in~\cite{felice2014information}.

\end{document}

%% file: macrosEl.tex
\newtheorem{remark}{Remark}
\newtheorem{example}{Example}
\newtheorem{proposition}{Proposition}
\newtheorem{property}{Property}
\newtheorem{corollary}{Corollary}
\newtheorem{definition}{Definition}
\newenvironment{proof}{\paragraph{Proof:}}{\hfill$\square$}

\def\supleft#1{{{}^{#1}}}
\def\TV{\mathrm{TV}}
\def\Eucl{\mathrm{Eucl}}
\def\Log{\mathrm{Log}}
\def\Exp{\mathrm{Exp}}
\def\dy{\mathrm{d}y}
\def\calH{\mathcal{H}}
\def\calA{\mathcal{A}}
\def\calB{\mathcal{B}}
\def\calC{\mathcal{C}}
\def\diag{\mathrm{diag}}
\def\calY{\mathcal{Y}}
\def\calG{\mathcal{G}}
\def\calX{\mathcal{X}}
\def\calC{\mathcal{C}}
\def\dnu{\mathrm{d}\nu}
\def\Length{\mathrm{Length}}
\def\FR{\mathrm{FR}}
\def\std{\mathrm{std}}
\def\barp{{\bar p}}
\def\calS{\mathcal{S}}
\def\mmax{{\mathrm{max}}}
\def\mmin{{\mathrm{min}}}

\def\dmu{\mathrm{d}\mu}
\def\Birkhoff{\mathrm{Birkhoff}}
\def\dx{\mathrm{d}x}
\def\bbP{\mathbb{P}}
\def\Fisher{\mathrm{Fisher}}
\def\dtheta{{\mathrm{d}\theta}}
\def\Jac{\mathrm{Jac}}
\def\KL{\mathrm{KL}}
\def\barP{{\bar{P}}}

\def\dS{\mathrm{d}S}
\def\Sym{\mathrm{Sym}}
\def\dSigma{\mathrm{d}\Sigma}
\def\calN{\mathcal{N}}

\def\Fisher{\mathrm{Fisher}}

\def\Wishart{\mathrm{Wishart}}
\def\arccos{\mathrm{arccos}}
\def\arccosh{\mathrm{arccosh}}
\def\Poincare{{\mbox{Poincar\'e}}}
\def\eps{\epsilon}
\def\barE{\bar{E}}
\def\SPD{\mathrm{SPD}}
\def\MTD{\mathrm{MTD}}
\def\MGGD{\mathrm{MGGD}}

\def\bbN{\mathbb{N}}

\def\dm{\mathrm{d}m}
\def\dV{\mathrm{d}V}

\def\dotm{\dot{m}}
\def\dotV{\dot{V}}

\def\tr{\mathrm{tr}}

\def\ddotm{\ddot{m}}

\def\ddotV{\ddot{V}}

\def\ds{\mathrm{d}s}

\def\calM{\mathcal{M}}
\def\Cov{\mathrm{Cov}}
\def\calB{\mathcal{B}}

\def\calN{\mathcal{N}}
\def\calC{\mathcal{C}}

\def\du{\mathrm{d}u}
\def\El{\mathrm{El}}
\def\det{\mathrm{det}}

\def\dv{\mathrm{d}v}
\def\dlambda{\mathrm{d}\lambda}
\def\dP{\mathrm{d}P}

\def\Hellinger{\mathrm{Hellinger}}
\def\length{\mathrm{length}}
\def\calB{\mathcal{B}}

\def\bbR{\mathbb{R}}
\def\dt{{\mathrm{d}t}}
\def\calE{\mathcal{E}}

\def\eps{\epsilon}

\def\dotV{{\dot{V}}}

\def\vectortwo#1#2{{\left[\begin{array}{l}#1 \cr #2 \end{array}\right]}}

\def\mattwotwo#1#2#3#4{{\left[\begin{array}{ll}#1 & #2 \cr  #3 & #4 \end{array}\right]}}

\def\dmu{\mathrm{d}\mu}

\def\st{{\ :\ }}